\documentclass[twocolumn,english,showpacs,pra]{revtex4-1}
\usepackage[T1]{fontenc}
\usepackage[latin9]{inputenc}
\usepackage{color}
\usepackage{amsmath}
\usepackage{amssymb}
\usepackage{graphicx}
\usepackage{esint}

\makeatletter
 \@ifundefined{textcolor}{}
 {%
   \definecolor{BLACK}{gray}{0}
   \definecolor{WHITE}{gray}{1}
   \definecolor{RED}{rgb}{1,0,0}
   \definecolor{GREEN}{rgb}{0,1,0}
   \definecolor{BLUE}{rgb}{0,0,1}
   \definecolor{CYAN}{cmyk}{1,0,0,0}
   \definecolor{MAGENTA}{cmyk}{0,1,0,0}
   \definecolor{YELLOW}{cmyk}{0,0,1,0}
 }

\usepackage{babel}
\usepackage{babel}
\usepackage{babel}

\usepackage{babel}

\makeatother

\usepackage{babel}
\begin{document}

\title{Role of thermal noise in tripartite quantum steering}

\author{Meng Wang$^{1}$, Qihuang Gong$^{1,2}$, Zbigniew Ficek$^{3}$, and
Qiongyi He$^{1,2}$}

\email{qiongyihe@pku.edu.cn}

\selectlanguage{english}%

\affiliation{$^{1}$State Key Laboratory of Mesoscopic Physics, School of Physics,
Peking University, Beijing 100871, P. R. China}

\affiliation{$^{2}$Collaborative Innovation Center of Quantum Matter, Beijing
100871, P. R. China}

\affiliation{$^{3}$National Centre for Applied Physics, KACST, P.O. Box 6086,
Riyadh 11442, Saudi Arabia}
\begin{abstract}
The influence of thermal noise on bipartite and tripartite quantum
steering induced by a short laser pulse in a hybrid three-mode optomechanical
system is investigated. The calculation is carried out under the bad
cavity limit, the adiabatic approximation of a slowly varying amplitude
of the cavity mode, and with the assumption of driving the cavity
mode with a blue detuned strong laser pulse. Under such conditions,
explicit expressions of the bipartite and tripartite steering parameters
are obtained, and the concept of collective tripartite quantum steering,
recently introduced by He and Reid {[}Phys. Rev. Lett. \textbf{111},
250403 (2013){]}, is clearly explored. It is found that both bipartite
and tripartite steering parameters are sensitive functions of the
initial state of the modes and distinctly different steering behaviour
could be observed depending on whether the modes were initially in
a thermal state or not. For the modes initially in a vacuum state,
the bipartite and tripartite steering occur simultaneously over the
entire interaction time. This indicates that collective tripartite
steering cannot be achieved. The collective steering can be achieved
for the modes initially prepared in a thermal state. We find that
the initial thermal noise is more effective in destroying the bipartite
rather than the tripartite steering which, on the other hand, can
persist even for a large thermal noise. For the initial vacuum state
of a steered mode, the tripartite steering exists over the entire
interaction time even if the steering modes are in very noisy thermal
states. When the steered mode is initially in a thermal state, it
can be collectively steered by the other modes. There are thresholds
for the average number of the thermal photons above which the existing
tripartite steering appears as the collective steering. Finally, we
point out that the collective steering may provide a resource in a
hybrid quantum network for quantum secret~sharing protocol. 
\end{abstract}


\maketitle

\section{Introduction}

The main property of entanglement is that it is shared equally between
the subsystems, that one cannot judge which of the subsystems is more
or less responsible for the entanglement \cite{werner}. Unlike entanglement,
quantum steering \cite{hw,eprappligrangcopy,jones,mdr13,ec13,mko13,qmNJP,cp14,bv14,sn14,cl14,sb14}
distinguishes the role of each subsystem. In fact, it is a form of
quantum nonlocality and gives a way to quantify how measurements by
Alice on her local particle $A$ can collapse the wavepacket of Bob's
particle $B$. The asymmetry reflects the asymmetric nature of the
original Einstein\textendash{}Podolsky\textendash{}Rosen (EPR) paradox
\cite{einstein,epr1989,rmp,Eric2009}, in which it is the reduced
uncertainty levels of Alice's predictions for Bob's system that are
relevant in establishing the paradox \cite{einstein,epr1989,rmp,oneway steering}.
Recently, the concept of collective multipartite steering has been
developed \cite{genuine steering}, which shows that special quantum
states allow Einstein's nonlocality to be shared among all observers
involved.

An experimental challenge is to observe the quantum nonlocality predicted
by EPR for macroscopic system. Optomechanical systems with nanomechanical
oscillators provide a natural setting for testing quantum nonlocality
of mesoscopic systems \cite{opto-ent,ag10,peter,HeReid2013pulseent,Hofer}.
The ability to cool optomechanical systems near their ground states
\cite{cooling} resulted in the demonstration of a number of quantum
effects such as quantum-state transfer \cite{zhang2003,exptransfer},
mechanical entanglement \cite{vg07,gv08,gb11,ab11,tgl11,am12,jl12,xz13,sa14},
mechanical squeezing \cite{tl13,aa14,wc14} and electromagnetically
induced transparency \cite{ha11,cs11,sn13}. In addition, spatial
entanglement between the macroscopic mirror and the microscopic cavity
field in a pulsed two-mode optomechanical system has been demonstrated
theoretically \cite{Hofer,HeReid2013pulseent}, and experimentally
observed \cite{exp_science2013}. Of particular interest is to observe
an EPR paradox for the position and momentum of mesoscopic mechanical
oscillators which could demonstrate the inconsistency of quantum mechanics
with the local reality of a macroscopic object.

When an optomechanical system is composed of more than two modes,
more complex correlations can be created. These correlations can significantly
affect the two-mode entanglement and result in multimode entanglement.
In searching for fully inseparable tripartite entanglement, first
extension is to study the entanglement between any pair of the tripartite
optomechanical systems \cite{2cavitymodes,gv08,ba11,cm11,sl12,xb12,rp12,wa13,ti13}.
Most of them produce a partially (at least one pair is entangled)
or fully inseparable (any two pairs show entanglement) tripartite
entanglement. It has been demonstrated that a genuine tripartite entanglement
can be produced in a three-mode optomechanical system composed of
an atomic ensemble located inside a single-mode cavity with a movable
mirror \cite{genu_ent_opto}.

In this paper, we examine the conditions for tripartite steering and
a special form of the tripartite steering called \textit{collective
tripartite quantum steering}~\cite{genuine steering}. We study a
hybrid optomechanical system composed of an ensemble of $N$ identical
two-level atoms located inside a single-mode cavity formed by two
mirrors, a fixed semitransparent mirror and a movable fully reflective
mirror. The cavity is driven by a short laser pulse and thus is free
from the restriction of the stability requirements. We adopt the criteria
for bipartite and tripartite steering determined by parameters which
are simple functions of the variances and correlation functions of
the quadrature components of the amplitudes of the output modes. We
are particularly interested in the role of the thermal noise in the
creation of collective tripartite steering between the modes. The
treatment is restricted to the bad cavity limit under which the adiabatic
approximation can be made of a slowly varying amplitude of the cavity
mode. In addition, the laser pulses are assumed to be strong and blue
detuned to the cavity and the atomic resonance frequencies. We show
that the thermal noise presented in the input modes is more effective
in destroying the bipartite rather than the tripartite steering which,
on the other hand, can persist even for a large thermal noise. The
threshold values for the average number of the thermal photons at
which the bipartite steering disappears are easy calculated.

The paper is organized as follows. In Sec. \ref{Sec2}, we introduce
parameters that determine the conditions for bipartite and tripartite
steering and briefly discuss the conditions required for the tripartite
steering to be regarded as collective tripartite steering. In Sec.~\ref{Sec3}
the pulsed three-mode optomechanics is introduced. We define a set
of normalized temporal modes and derive analytic expressions for the
variances and correlation function of the quadrature components of
the output fields. In Sec.~\ref{Sec4}, we evaluate the parameters
for the bipartite and tripartite steering and discuss in details the
conditions for collective tripartite steering to occur in the system.
We summarize our results in Sec.~\ref{Sec5}. Finally, in the Appendix,
we give general expressions for the bipartite and tripartite steering
parameters in terms of the variances and correlation functions, and
optimal weight factors that minimize the variances involved in the
steering parameters.

\section{Definitions and identification of collective tripartite steering}

\label{Sec2}

For later convenience we start by introducing the definition of tripartite
steering and explain in details how one could distinguish between
the ordinary tripartite steering and collective tripartite steering.
We introduce parameters that measures the degree of ordinary and collective
tripartite steering.

Quantum steering is normally identified by criteria which are a natural
generalization of those for entanglement. They involve inequalities
the variances must satisfy which, in fact, are stronger than those
for entanglement~\cite{rmp}. Therefore, steering always certifies
entanglement. Let us briefly discuss the criteria for bipartite
and tripartite steering. The criteria are based on an accuracy of
inference defined as the root mean square of the variances $\Delta_{inf,j}^{2}X_{i}^{{\rm out}}$ and $\Delta_{inf,j}^{2}P_{i}^{{\rm out}}$
of the conditional distributions $P(X_{i}^{{\rm out}}|O_{j}^{{\rm out}})$ and $P(P_{i}^{{\rm out}}|O_{j}^{{\rm out\prime}})$
($O_{j}^{{\rm out}}, O_{j}^{{\rm out}\prime}\equiv X_{j}^{{\rm out}},P_{j}^{{\rm out}})$,
for a result of measurement of the quadratures $X_{i}^{{\rm out}}, P_{i}^{{\rm out}}$
at $i$, based on the results $O_{j}^{{\rm out}}, O_{j}^{{\rm out}\prime}$ of the measurement
at $j$ \cite{epr1989,rmp}. Here,  $O_{j}^{{\rm out}}, O_{j}^{{\rm out}\prime}$ are arbitrary observables (quadratures) for system $j$ selected such that they {\it minimize} the variance product, $\Delta_{inf,j}^{2}X_{i}^{{\rm out}}\Delta_{inf,j}^{2}P_{i}^{{\rm out}}$.
A useful strategy is to use a linear estimate $u_{j}O_{j}^{{\rm out}}$, where $u_{j}$ is a constant
chosen such that it minimizes the variance product \cite{epr1989,rmp}. The inferred uncertainty
$\Delta_{inf,j}X_{i}^{{\rm out}}$ can be written as 
\begin{align}
\Delta_{inf,j}X_{i}^{{\rm out}}=\Delta(X_{i}^{{\rm out}}+u_{j}O_{j}^{{\rm out}}), \label{equ1}
\end{align}
where the quadrature $O_{j}$ is selected either $O_{j}\equiv X_{j}$
or $O_{j}\equiv P_{j}$, depending on the type of the correlations
between the modes $i$ and $j$ \cite{rmp,Eric2009}, and we use the
notation $\Delta X\equiv\sqrt{\langle X^{2}\rangle-\langle X\rangle^{2}}$. The best choice of $u_{j}$ can be calculated by linear regression and it is not difficult to show that the uncertainty (\ref{equ1}) is minimized for $u_{j}=-\langle X_{i}^{{\rm out}},O_{j}^{{\rm out}}\rangle/\Delta^{2} O_{j}^{{\rm out}}$.

We say that the mode $i$ is steered by the mode $j$ if the product
of the inferred variances satisfies the inequality ($\hbar=1$) \cite{epr1989,Eric2009}
\begin{equation}
E_{i|j}=\Delta_{inf,j}X_{i}\Delta_{inf,j}P_{i}<\frac{1}{2}.\label{bi-steering}
\end{equation}

The condition for tripartite steering is described in terms of the
inferred variances of a linear combination of the quadrature components
\begin{equation}
\Delta_{inf,jk}X_{i}^{{\rm out}}=\Delta\left[X_{i}^{{\rm out}}+\left(u_{j}O_{j}^{{\rm out}}+u_{k}O_{k}^{{\rm out}}\right)\right],\label{eq3}
\end{equation}
where, depending on the type of correlations between the modes, the
quadrature $O_{j(k)}^{{\rm out}}$ can be selected either $X_{j(k)}^{{\rm out}}$
or $P_{j(k)}^{{\rm out}}$, and the weight factors $u_{j},u_{k}$
are estimated to minimize the variance. $\Delta_{inf,jk}P_{i}^{{\rm out}}$
is defined similarly. Then, we say that the mode $i$ is steered by
the group of modes $\{jk\}$ if 
\begin{equation}
E_{i|jk}=\Delta_{inf,jk}X_{i}^{{\rm out}}\Delta_{inf,jk}P_{i}^{{\rm out}}<\frac{1}{2}.\label{eq:groupsteer}
\end{equation}
It involves a superposition of the modes $j$ and $k$ which can be
treated a single ``collective'' mode $W_{jk}^{{\rm out}}=O_{j}^{{\rm out}}+u_{jk}O_{k}^{{\rm out}}$.
Thus, the collective mode $W_{jk}^{{\rm out}}$ can be treated as
a single mode that can steer the mode $i$.

The inequality (\ref{eq:groupsteer}) is the sufficient condition
for tripartite steering without any requirements about the bipartite
steering between modes $i$ and $j$, and between $i$ and~$k$.
This means that in general for a tripartite steering we can have two
distinct possibilities. Namely, we could have $E_{i|jk}<1/2$ with
both or either $E_{i|j}$ or $E_{i|k}$ smaller than $1/2$. In this
case, the tripartite steering is accompanied by a bipartite steering,
and is referred to as ordinary tripartite steering. The other case
corresponds to the inequality $E_{i|jk}<1/2$ with both $E_{i|j}\geq1/2$
and $E_{i|k}\geq1/2$. In this case, the tripartite steering is not
accompanied by the bipartite steering. The mode $i$ is steered solely
by the collective mode and therefore it is referred to as collective
steering~\cite{genuine steering}. In other words, to demonstrate
the existence of collective steering in a tripartite system we must
show that whenever the condition $E_{i|jk}<1/2$ holds, the bipartite
steering parameters are $E_{i|j}\geq1/2$ and $E_{i|k}\geq1/2$. The
collective steering is thus a generalization of the ordinary tripartite
steering to the case when a given mode is steered only by the collective
mode of a linear superposition of the remaining modes.

\section{Hybrid pulsed cavity optomechanical system}

\label{Sec3}

We now illustrate how the ordinary and collective tripartite steering
may be created in a three mode system. We choose a three-mode hybrid
pulsed optomechanical system and investigate under which conditions
the ordinary tripartite steering can be created and under what circumstances
it is not accompanied by the bipartite steering. The three-mode hybrid
pulsed optomechanical system is known to exhibit bipartite steering~\cite{genu_ent_opto}.

\begin{figure}[h]
\centering{}\includegraphics[width=0.9\columnwidth]{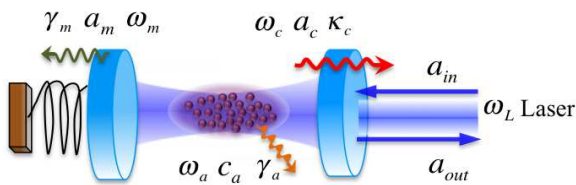} \caption{(Color online) Schematic diagram of a driven hybrid optomechanical
system. The cavity mode, atoms, and movable mirror constitute a three-mode
system and are represented by annihilation operators $a_{c},c_{a}$
and $a_{m}$, respectively. Here, $a_{in}$ and $a_{out}$ denote
input and output cavity fields.\label{fig:1} }
\end{figure}

We consider an optomechanical system composed of a single-mode cavity
with a movable fully reflective mirror, containing an atomic ensemble
and driven by light pulses of duration $\tau$, as shown in Fig. \ref{fig:1}.
The cavity mode has a frequency $\omega_{c}$ and a decay rate $\kappa$.
The atomic ensemble contains $N$ identical two-level atoms each composed
of a ground state $|1_{j}\rangle$ and an excited state $|2_{j}\rangle$
($j=1,\cdots N$), separated by the transition frequency $\omega_{a}$.
We represent the atomic ensemble in terms of the collective dipole
lowering $S^{-}=\sum_{j}|1_{j}\rangle\langle2_{j}|$, raising $S^{+}=\sum_{j}|2_{j}\rangle\langle1_{j}|$
and population inversion $S_{z}=\sum_{j}(|2_{j}\rangle\langle2_{j}|-|1_{j}\rangle\langle1_{j}|)$
operators. We assume that the atomic ensemble is composed of a large
number of atoms $(N\gg1)$ which allows us to make use of the Holstein-Primakoff
representation~\cite{s21} that transforms the collective atomic
operators into bosonic annihilation and creation operators $c_{a}$
and $c_{a}^{\dagger}$: 
\begin{eqnarray}
S^{+}=(S^{-})^{\dagger}=c_{a}^{\dag}\sqrt{N-c_{a}^{\dag}c_{a}},\quad S_{z}=c_{a}^{\dag}c_{a}-N.\label{e6}
\end{eqnarray}
If the atomic ensemble is weakly coupled to the cavity mode the mean
number of atoms transferred to the upper state $|2_{j}\rangle$ is
expected to be much smaller than the total number of atoms, i.e.,
$\langle c_{a}^{\dag}c_{a}\rangle\ll N$. By expanding the square
root in Eq.~(\ref{e6}) and neglecting terms of the order of $\emph{O}(1/N)$,
the collective atomic operators can be approximated as~\cite{s20}
\begin{eqnarray}
S^{+}\approx\sqrt{N}c_{a}^{\dagger},\quad S^{-}\approx\sqrt{N}c_{a},\quad S_{z}\approx\langle S_{z}\rangle\approx-N.
\end{eqnarray}
It is easily verified that the operators $c_{a}$ and $c_{a}^{\dagger}$
satisfy the fundamental commutation relation for boson operators,
$[c_{a},c_{a}^{\dagger}]=1$.

The Hamiltonian of the system, in a frame rotating with the laser
frequency $\omega_{L}$, is given by \cite{genu_ent_opto} 
\begin{eqnarray}
H & = & \hbar\Delta_{c}a_{c}^{\dagger}a_{c}+\hbar\omega_{m}a_{m}^{\dagger}a_{m}+\hbar\Delta_{a}c_{a}^{\dagger}c_{a}\nonumber \\
 &  & +\hbar g_{0}a_{c}^{\dagger}a_{c}\left(a_{m}^{\dagger}+a_{m}\right)+\hbar g_{a}\left(c_{a}^{\dagger}a_{c}+a_{c}^{\dagger}c_{a}\right)\nonumber \\
 &  & +i\hbar\left[E(t)a_{c}^{\dagger}-E^{\ast}(t)a_{c}\right].\label{q2}
\end{eqnarray}
The first three terms represent the energy of the modes. Here, $\Delta_{c}=\omega_{c}-\omega_{L}$
and $\Delta_{a}=\omega_{a}-\omega_{L}$ are the detunings of the laser
frequency $\omega_{L}$ from the cavity and the atomic transition
frequencies, respectively. The forth term describes the interaction
of the cavity mode with the movable mirror. This is a nonlinear type
interaction with the strength determined by single-photon coupling
constant $g_{0}$. As we shall see this interaction results in a parametric
coupling between the modes when the cavity is driven by a blue detunned
laser \cite{Vitali2008}. The fifth term describes the interaction
between cavity mode and atomic excitation mode with coupling constant
$g_{a}$. This is a beamsplitter-like interaction. The last term in
Eq.~(\ref{q2}) describes the interaction of the cavity mode with
the coherent laser field of the amplitude $E(t)$. The laser field
is injected into the cavity mode through the fixed mirror. Note that
the atomic mode is not directly coupled to the mechanical mode.

The evolution of the system is studied using the Heisenberg equations
of motion~\cite{Hofer,genu_ent_opto}. The equations form a set of
coupled nonlinear differential equations, which we solve in the limit
of a strong pulse, $|E(t)|\gg g_{0},g_{a}$. In this case, we can
make the semi-classical approximation in which we write the operators
of modes as composed of a large classical amplitude and a small fluctuation
operator, i.e. $a_{i}\rightarrow\alpha_{i}+\delta a_{i}$ ($i=c,m$)
and $c_{a}\rightarrow\bar{c}_{a}+\delta c_{a}$. The fluctuation operators,
in a frame rotating with $\omega_{m}$ and under the rotating-wave
approximation in which we ignore all terms oscillating with $2\omega_{m}$,
satisfy the following linearized quantum Langevin equations 
\begin{align}
\delta\dot{a}_{m}= & -\gamma_{m}\delta a_{m}-ig\delta a_{c}^{\dagger}-\sqrt{2\gamma_{m}}\,\xi_{{\rm in}}^{m},\nonumber \\
\delta\dot{a}_{c}= & -\kappa\delta a_{c}-ig_{a}\delta a_{a}-ig\delta a_{m}^{\dagger}-\sqrt{2\kappa}\,\xi_{{\rm in}}^{a},\nonumber \\
\delta\dot{c}_{a}= & -\gamma_{a}\delta c_{a}-ig_{a}\delta a_{c}-\sqrt{2\gamma_{a}}\,\xi_{{\rm in}}^{c},\label{q12}
\end{align}
where $g=g_{0}|\alpha_{c}|$ is the effective optomechanical coupling
constant and we have assumed that the atomic detuning $\Delta_{a}=-\omega_{m}$
and the effective cavity detuning $\Delta_{c}+g_{0}(\alpha_{m}+\alpha_{m}^{\ast})=-\omega_{m}$.
The choice we have made for the detunings corresponds to the laser
pulse driving on the blue sideband of the cavity and the atomic resonances.

The fluctuation operators are affected by input noises~$\xi_{{\rm in}}^{i}$
arising from the coupling of the modes to their surrounding environments.
Because it is precisely the effect of the noise on the dynamics of
the modes that interests us most here, we assume that the environments
are in thermal vacuum states characterized by the correlation functions
$\langle\xi_{{\rm in}}^{i}(t)\xi_{{\rm in}}^{i}(t^{\prime})+\xi_{{\rm in}}^{i}(t^{\prime})\xi_{{\rm in}}^{i}(t)\rangle=(2n_{i}+1)\delta(t-t')$,
where $n_{i}$ is the average number of thermal photons in the environment
coupled to the mode.

\subsection{Normalized temporal modes}

In the bad cavity limit $\kappa\gg g_{a},g$ and for short evolution
times $t\sim1/\kappa$, we may neglect the relaxations of the atoms
and the mechanical mirror $(\gamma_{a}=\gamma_{m}=0)$. In this case,
Eq.~(\ref{q12}) are simple enough to be solved analytically. In particular,
when we make an adiabatic approximation, $\delta\dot{a}_{c}\approx0$,
which is justified for $\kappa\gg g_{a},g$, we arrive at the following
equations 
\begin{align}
a_{c}(t) & \approx-i\frac{g_{a}}{\kappa}c_{a}(t)-i\frac{g}{\kappa}a_{m}^{\dagger}(t)-\sqrt{\frac{2}{\kappa}}a_{{\rm in}}(t),\label{q16}\\
a_{m}(t) & =a_{m}(0){\rm e}^{Gt}+\sqrt{GG_{a}}{\rm e}^{Gt}\int_{0}^{t}dt^{\prime}c_{a}^{\dagger}(t^{\prime}){\rm e}^{-Gt^{\prime}}\nonumber \\
 & +i\sqrt{2G}{\rm e}^{Gt}\int_{0}^{t}dt^{\prime}a_{{\rm in}}^{\dagger}(t^{\prime}){\rm e}^{-Gt},\label{q18}\\
c_{a}^{\dagger}(t) & =c_{a}^{\dagger}(0){\rm e}^{-G_{a}t}-\sqrt{GG_{a}}{\rm e}^{-G_{a}t}\int_{0}^{t}dt^{\prime}a_{m}(t^{\prime}){\rm e}^{G_{a}t^{\prime}}\nonumber \\
 & -i\sqrt{2G_{a}}{\rm e}^{-G_{a}t}\int_{0}^{t}dt^{\prime}a_{{\rm in}}^{\dagger}(t^{\prime}){\rm e}^{G_{a}t^{\prime}}.\label{q19}
\end{align}
where $G=g^{2}/\kappa$ and $G_{a}=g_{a}^{2}/\kappa$. Note that for
simplicity of the notation, we have dropped $\delta$.

The structure of the solutions given by Eqs.~(\ref{q16})-(\ref{q19})
suggests the introduction of normalized temporal modes of the input
and output cavity fields, which in the case of $G>G_{a}$ are defined
by 
\begin{align}
A_{{\rm in}} & =\sqrt{\frac{2(G-G_{a})}{1-{\rm e}^{-2(G-G_{a})\tau}}}\int_{0}^{\tau}dt\, a_{{\rm in}}(t){\rm e}^{-(G-G_{a})t},\nonumber \\
A_{{\rm out}} & =\sqrt{\frac{2(G-G_{a})}{{\rm e}^{2(G-G_{a})\tau}-1}}\int_{0}^{\tau}dt\, a_{{\rm out}}^{c}(t){\rm e}^{(G-G_{a})t},\label{eq:q35}
\end{align}
where $a_{{\rm out}}^{c}(t)$ is the annihilation operator of the
output cavity field, given by the standard cavity input-output relation,
$a_{{\rm out}}^{c}(t)=a_{{\rm in}}(t)+\sqrt{2\kappa}a_{c}(t)$ \cite{gc85},
and $\tau$ is the duration of the laser pulse. We may also define
the normalized input and output operators of the atomic and the mirror
modes, $B_{{\rm in}}=a_{m}(0),\ B_{{\rm out}}=a_{m}(\tau),\ C_{{\rm in}}=c_{a}(0),\ C_{{\rm out}}=c_{a}(\tau)$,
and then find using Eqs.~(\ref{q16})-(\ref{q19}) that the solution
for the quadrature components $X_{i}$ and $P_{i}$ of the output
fields can be expressed only in terms of the quadrature components
of the input fields. Assuming that the input modes are in thermal
states characterized by the variances 
\begin{align}
\Delta^{2}X_{m}^{in}=\Delta^{2}P_{m}^{in} & =\left(n_{0}+\frac{1}{2}\right),\nonumber \\
\Delta^{2}X_{a}^{in}=\Delta^{2}P_{a}^{in} & =\left(n_{1}+\frac{1}{2}\right),\nonumber \\
\Delta^{2}X_{c}^{in}=\Delta^{2}P_{c}^{in} & =\left(n_{1}+\frac{1}{2}\right),
\end{align}
in which $n_{0}$ is the average numbers of thermal photons in the
mirror mode $m$ and $n_{1}$ is the average number of thermal photons
in the cavity mode $a$ and atomic mode $c$, we arrive to the following
expressions for the variances of the output fields 
\begin{align}
\Delta^{2}X_{a} & =\Delta^{2}P_{a}=\left(n_{1}+\frac{1}{2}\right)\nonumber \\
 & +\left(n_{0}+n_{1}+1\right)\alpha^{2}\left(e^{2r_{\alpha}}-1\right),\nonumber \\
\Delta^{2}X_{c} & =\Delta^{2}P_{c}=\left(n_{1}+\frac{1}{2}\right)\nonumber \\
 & +\left(n_{0}+n_{1}+1\right)\alpha^{2}\beta^{2}\left(e^{r_{\alpha}}-1\right)^{2},\nonumber \\
\Delta^{2}X_{m} & =\Delta^{2}P_{m}=\left(n_{0}+\frac{1}{2}\right)\nonumber \\
 & +\left(n_{0}+n_{1}+1\right)\left[\left(\alpha^{2}e^{r_{\alpha}}-\beta^{2}\right)^{2}-1\right],\nonumber \\
\left\langle X_{m},P_{a}\right\rangle  & =\left\langle P_{m},X_{a}\right\rangle \nonumber \\
 & =-\left(n_{0}+n_{1}+1\right)\alpha\sqrt{e^{2r_{\alpha}}-1}\left(\alpha^{2}e^{r_{\alpha}}-\beta^{2}\right),\nonumber \\
\left\langle X_{m},X_{c}\right\rangle  & =-\left\langle P_{m},P_{c}\right\rangle \nonumber \\
 & =-\left(n_{0}+n_{1}\!+\!1\right)\alpha\beta\!\left(e^{r_{\alpha}}-1\right)\!\left(\alpha^{2}e^{r_{\alpha}}-\beta^{2}\right),\nonumber \\
\left\langle P_{a},X_{c}\right\rangle  & =-\left\langle X_{a},P_{c}\right\rangle \nonumber \\
 & =\left(n_{0}+n_{1}+1\right)\alpha^{2}\beta\sqrt{e^{2r_{\alpha}}-1}\left(e^{r_{\alpha}}-1\right),\nonumber \\
\left\langle P_{m},X_{c}\right\rangle  & =\left\langle X_{c},P_{m}\right\rangle \!=\!\left\langle P_{c},X_{m}\right\rangle \!=\!\left\langle X_{m},P_{c}\right\rangle \!=\!0.\label{eq17}
\end{align}
where $r_{\alpha}=(G-G_{a})\tau=G\tau/\alpha^{2}=r/\alpha^{2}$ is
the normalized interaction time parameter, $\alpha=\sqrt{G/(G-G_{a})}$,
and $\beta=\sqrt{G_{a}/(G-G_{a})}$. Note that the parameter $r_{\alpha}$
has the physical meaning of the squeezing parameter~\cite{Hofer}.

The solutions for the variances of the output modes are used in the
following section to analyze the criteria for tripartite steering
and collective tripartite steering. We address the question of the
role of the thermal noise in the creation of ordinary tripartite steering
and collective tripartite steering.

\section{Tripartite steering and collective steering}

\label{Sec4}

We now proceed to evaluate the parameters $E_{i|jk}$ for different
combinations of the three modes of the optomechanical system to determine
the role of the inter-modal interactions and thermal noise in the
creation of tripartite steering. Our objective is to find simple analytic
forms for the steering parameters. We examine separately several cases
of different initial (input) states of the modes, the vacuum state
with $n_{0}=n_{1}=0$, thermal states with equal $(n_{0}=n_{1})$
and different $(n_{0}\neq n_{1})$ average numbers of thermal photons.
Next, we turn to the problem of tripartite collective steering which
requires the absence of bipartite steering between the modes. In trying
to produce such conditions, we note the constructive role of the thermal
noise in the collective steering of the modes.

\subsection{Tripartite steering}

\label{sec:4a}

For the three-mode system considered here, each mode can be steered
by the remaining two modes. In this case we have three combinations
for the steering parameter. In the first combination, $E_{m|ac}$
describes steering of the mirror mode $m$ by the cavity and atomic
modes, $E_{c|am}$ describes steering of the atomic mode by the cavity
and mirror modes, and $E_{a|mc}$ describes steering of the cavity
mode by the mirror and atomic modes.

To evaluate the tripartite steering parameters $E_{i|jk}$, we take
suitable linear combinations of the quadrature components and find
\begin{align}
E_{m|ac} & =\Delta\!\left[X_{m}\!+\!\left(u_{a}P_{a}\!+\! u_{c}X_{c}\right)\right]\!\Delta\!\left[P_{m}\!+\!\left(u_{a}X_{a}\!-\! u_{c}P_{c}\right)\right]\nonumber \\
 & =\Delta^{2}X_{m}+u_{a}^{2}\Delta^{2}P_{a}+u_{c}^{2}\Delta^{2}X_{c}+2u_{a}\left\langle X_{m},P_{a}\right\rangle \nonumber \\
 & \ \ \ \ \ \ \ \ +2u_{c}\left\langle X_{m},X_{c}\right\rangle +2u_{a}u_{c}\left\langle P_{a},X_{c}\right\rangle ,\label{eq15}\\
E_{c|am} & =\Delta\!\left[X_{c}\!+\!\left(u_{a}P_{a}\!+\! u_{m}X_{m}\right)\right]\!\Delta\!\left[P_{c}\!-\!\left(u_{a}X_{a}\!+\! u_{m}P_{m}\right)\right]\nonumber \\
 & =\Delta^{2}X_{c}+u_{a}^{2}\Delta^{2}P_{a}+u_{m}^{2}\Delta^{2}X_{m}+2u_{a}\left\langle X_{c},P_{a}\right\rangle \nonumber \\
 & \ \ \ \ \ \ \ \ +2u_{m}\left\langle X_{c},X_{m}\right\rangle +2u_{a}u_{m}\left\langle P_{a},X_{m}\right\rangle ,\label{eq15a}\\
E_{a|mc} & =\Delta\!\left[X_{a}\!+\!\left(u_{m}P_{m}\!+\! u_{c}P_{c}\right)\right]\!\Delta\!\left[P_{a}\!+\!\left(u_{m}X_{m}\!-\! u_{c}X_{c}\right)\right]\nonumber \\
 & =\Delta^{2}X_{a}+u_{m}^{2}\Delta^{2}P_{m}+u_{c}^{2}\Delta^{2}P_{c}+2u_{m}\left\langle X_{a},P_{m}\right\rangle \nonumber \\
 & \ \ \ \ \ \ \ \ +2u_{c}\left\langle X_{a},P_{c}\right\rangle +2u_{m}u_{c}\left\langle P_{m},P_{c}\right\rangle .\label{eq15b}
\end{align}
This shows that, apart from the variances of the quadratures involved,
the parameters depend on correlations between the modes. Since the
variances $\Delta^{2}X_{i}\geq1/2$ and $\Delta^{2}P_{i}\geq1/2$,
we see that the mechanism for steering is in the correlations between
the modes. Steering will occur if the correlations are sufficiently
large and negative to enforce the inequality $E_{i|jk}<1/2$. The
minimum requirement for this to be possible is that there are negative
correlations at least between two modes. If the modes are uncorrelated,
$\left\langle X_{i},P_{j}\right\rangle =\left\langle X_{i},X_{j}\right\rangle =\left\langle P_{i},X_{j}\right\rangle =0$,
and then all the steering parameters are greater than $1/2$. Therefore,
correlations between the modes are necessary to produce quantum steering.
Note that the requirement that the modes should be correlated is necessary
but not sufficient for steering. The correlations may be negative
but not large enough to reduce the steering parameter below the threshold
for steering.

The steering parameters $E_{i|jk}$ involve correlations between the
steering modes $j$ and $k$. It is well known that negative correlations
is one mechanism for entanglement between modes. Therefore, we would
expect an enhancement of steering when $j$ and $k$ are entangled.
However, we will see that a better tripartite steering is obtained
when the correlations between the steering modes are positive rather
than negative.

To see if tripartite steering exists in the system and especially
what is the role of the input thermal noise, we evaluate the steering
parameters given by the expressions~(\ref{eq15}-\ref{eq15b}). The
general solutions, Eq.~(\ref{eq17}), for the variances and the correlation
functions involved in these expressions are simple enough to obtain
the analytical expressions for the steering parameters. After straightforward
but somewhat tedious manipulation of terms, we arrive at the following
general solutions 
\begin{align}
E_{m|ac} & =\left(n_{0}+\frac{1}{2}\right)\nonumber \\
 & \times\left\{ 1-\frac{\left(2\bar{n}+1\right)\!\left[\left(\alpha^{2}e^{r_{\alpha}}\!-\!\beta^{2}\right)^{2}-1\right]}{\left(n_{1}+\frac{1}{2}\right)\!+\!\left(2\bar{n}+1\right)\!\left[\left(\alpha^{2}e^{r_{\alpha}}\!-\!\beta^{2}\right)^{2}\!-\!1\right]}\!\right\} ,\nonumber \\
E_{c|am} & =\left(\! n_{1}\!+\!\frac{1}{2}\right)\!\!\left[1-\frac{\alpha^{2}\beta^{2}\!\left(2\bar{n}+1\right)\!\left(e^{r_{\alpha}}\!-\!1\right)^{2}}{\left(n_{0}\!+\!\frac{1}{2}\right)\!+\!\alpha^{2}\beta^{2}\!\left(2\bar{n}\!+\!1\right)\!\left(e^{r_{\alpha}}\!-\!1\right)^{2}}\!\right],\nonumber \\
E_{a|mc} & =\left(\! n_{1}\!+\!\frac{1}{2}\right)\!\!\left[1-\frac{\alpha^{2}\left(2\bar{n}+1\right)\!\left(e^{2r_{\alpha}}-1\right)}{\left(n_{0}\!+\!\frac{1}{2}\right)\!+\!\alpha^{2}\!\left(2\bar{n}\!+\!1\right)\!\left(e^{2r_{\alpha}}\!-\!1\right)}\right],\label{eq19p}
\end{align}
in which $2\bar{n}=(n_{0}+n_{1})$. 

It is seen from Eq.~(\ref{eq19p}) that in the absence of the thermal
noise $(n_{0}=n_{1}=0)$ all the parameters are smaller than $1/2$
indicating that a tripartite steering of each mode occurs immediately
when the laser pulse is turned on, $r_{\alpha}>0$. In the presence
of the thermal noise the parameters are enhanced and thermal barriers
appear that the tripartite steering of a given mode occurs at a finite
$r_{\alpha}$. It is interesting that the thermal barriers are determined
by the thermal noise present at the steered mode only. For example,
the parameter $E_{m|ac}$ is enhanced by the factor $(n_{0}+1/2)$,
the thermal noise at the steered mode $m$. Similarly, the parameters
$E_{c|am}$ and $E_{a|mc}$ are enhanced by the factor $(n_{1}+1/2)$,
the thermal noise at the steered modes $c$ and $a$.

The steering parameters increase with the thermal noise but remain
smaller than $1/2$ at least for some maximum (threshold) values of
$n_{0}$ and $n{}_{1}$ determined by $r_{\alpha}$ and $\alpha$.
For example, in the case when the modes are equally affected by the
thermal noise, $n_{0}=n_{1}=n$, the parameters can be rewritten as
\begin{align}
E_{m|ac} & =\frac{1}{2}+\frac{n-\alpha^{2}\!\left(e^{r_{\alpha}}\!-\!1\right)\!\left(\alpha^{2}e^{r_{\alpha}}\!-\!\beta^{2}\!+\!1\right)}{2\left[\alpha^{2}\left(e^{r_{\alpha}}-1\right)+1\right]^{2}-1},\nonumber \\
E_{c|am} & =\frac{1}{2}+\frac{n-\alpha^{2}\beta^{2}\left(e^{r_{\alpha}}-1\right)^{2}}{2\alpha^{2}\beta^{2}\left(e^{r_{\alpha}}-1\right)^{2}+1},\nonumber \\
E_{a|mc} & =\frac{1}{2}+\frac{n-\alpha^{2}\!\left(e^{2r_{\alpha}}-1\right)}{2\alpha^{2}\left(e^{2r_{\alpha}}-1\right)+1}.\label{eq19}
\end{align}
The threshold values of $n$ at which the tripartite steering of the
modes disappears are given by 
\begin{align}
n_{{\rm th}} & =\alpha^{2}\!\left(e^{r_{\alpha}}\!-\!1\right)\!\left(\alpha^{2}e^{r_{\alpha}}\!-\!\beta^{2}\!+\!1\right),\quad{\rm for}\ E_{m|ac},\nonumber \\
n_{{\rm th}} & =\alpha^{2}\beta^{2}\left(e^{r_{\alpha}}-1\right)^{2},\quad{\rm for}\ E_{c|am},\nonumber \\
n_{{\rm th}} & =\alpha^{2}\!\left(e^{2r_{\alpha}}-1\right),\quad{\rm for}\ E_{a|mc}.\label{eq19u}
\end{align}
Note that the threshold values increase exponentially with $r_{\alpha}$,
thus tripartite steering can be preserved even in the presence of
a large thermal noise.

\begin{figure}[h]
\begin{centering}
\includegraphics[width=0.95\columnwidth]{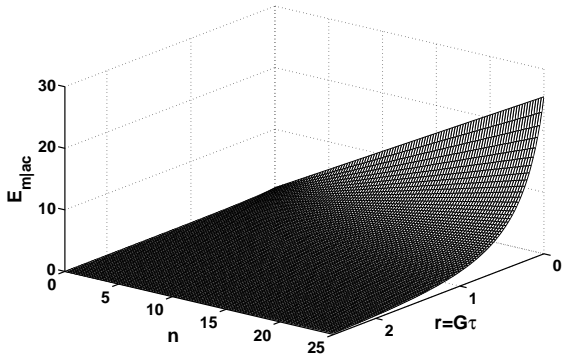} 
\par\end{centering}

\caption{The tripartite steering parameter $E_{m|ac}$ plotted as a function
of $r=G\tau=r_{\alpha}\alpha^{2}$ and $n$ for the case with $\alpha=1.2$.
\label{fig:2} }
\end{figure}

Viewed as a function of $r_{\alpha}$, the tripartite steering of
the modes appears at a finite $r_{\alpha}$. For example, for the
case of the mode $m$, steering occurs at 
\begin{eqnarray}
r_{\alpha}=\ln\left(1+\frac{\sqrt{n+1}-1}{\alpha^{2}}\right),\label{eq19c}
\end{eqnarray}
which is different from zero when $n\neq0$. The threshold value of
$r_{\alpha}$ increases with $n$. This shows that a larger thermal
noise requires a larger squeezing to produce a tripartite steering.

The above considerations are illustrated in Fig.~\ref{fig:2}. In
the absence of thermal noise $(n=0)$ the tripartite steering is present
over the entire range of $r$ and perfect steering, $E_{m|ac}=0$,
is achieved for $r\rightarrow\infty$. In the presence of thermal
noise $(n\neq0)$, there is a threshold for $r$ above which the steering
takes place. 

\begin{figure}[h]
\begin{centering}
\includegraphics[width=0.95\columnwidth]{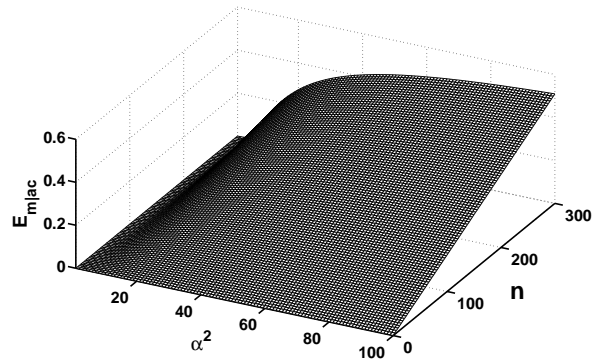} 
\par\end{centering}

\caption{Tripartite steering parameter $E_{m|ac}$ as a function of $\alpha^{2}$
and $n$ for $r=15$. \label{fig:3} }
\end{figure}

Although the steering parameters (\ref{eq19}) go up with an increasing
$n$, it does not prevent us from achieving perfect steering of the
modes. It is easily verified from Eq.~(\ref{eq19}) that in the limit
of large squeezing, $r\gg1$, the steering parameters reduce to simple
expressions 
\begin{align}
E_{m|ac} & \approx\frac{n+1/2}{2\alpha^{4}e^{2r_{\alpha}}},\nonumber \\
E_{c|am} & \approx\frac{n+1/2}{2\alpha^{2}\beta^{2}e^{2r_{\alpha}}},\nonumber \\
E_{a|mc} & \approx\frac{n+1/2}{2\alpha^{2}e^{2r_{\alpha}}}.\label{eq19k}
\end{align}
We see here that, even when the thermal noise is large, the steering
parameters can be made negligibly small by increasing the squeezing
parameter $r_{\alpha}$. This shows that the effect of the thermal
noise on the tripartite steering is not dramatic and perfect steering
can be observed even for large $n$.

Figure~\ref{fig:3} shows the variation of the parameter $E_{m|ac}$
with $\alpha^{2}$ and $n$ for $r=15$. Note that $\alpha^{2}$ depends
on the relative strength of the coupling constants $G$ and $G_{a}$
between the modes. For small $\alpha^{2}$, corresponding to the parametric
type interaction dominating over the beamsplitter type interaction,
we see that perfect steering can be observed over the entire range
of~$n$. This tendency continues until at $\alpha^{2}\approx2$ the
degree of steering starts to decrease and becomes independent of $\alpha^{2}$
for $\alpha^{2}\gg1$. A considerable tripartite steering still is
present even for large $n$. In order to see it explicitly, we take
the limit of $\alpha^{2}\gg r$ and expand the exponents appearing
in the expressions (\ref{eq19}) into Taylor series and obtain 
\begin{align}
E_{m|ac} & \approx\frac{n+1/2}{2(r+1)^{2}-1},\nonumber \\
E_{c|am} & \approx\frac{n+1/2}{2\beta^{2}r^{2}/\alpha^{2}+1},\nonumber \\
E_{a|mc} & \approx\frac{n+1/2}{4r+1}.\label{eq19e}
\end{align}

We see that the steering parameter $E_{m|ac}$ are independent of
$\alpha^{2}$. What this means is that the tripartite steering can
be present over a large range of $n$ even if the beamsplitter type
interaction, determined by $G_{a}$, is comparable to the parametric
type interaction, determined by $G$.

Let us now comment about the dependence of the steering parameters
on the sign of the correlations between the steering modes. In steering
of the mode $m$ by the pair $\{ac\}$, Eq.~(\ref{eq15}), the correlation
between the steering modes is described by the correlation function
$\langle P_{a},X_{c}\rangle$. In steering of the mode $c$, Eq.~(\ref{eq15a}),
the correlation between the steering modes is described by $\langle P_{a},X_{m}\rangle$.
According to Eq.~(\ref{eq17}), $\langle P_{a},X_{c}\rangle$ is
positive whereas $\langle P_{a},X_{m}\rangle$ is negative. This suggests
that the involvement of the negative correlation should result in
a better steering of the mode $c$ by the pair $\{am\}$ than the
mode $m$ by the pair $\{ac\}$. However, this is not the case, a
negative correlation between the steering modes not necessarily leads
to a better steering. To demonstrate this feature we take the ratio
$E_{c|am}/E_{m|ac}$ and find 
\begin{align}
\frac{E_{c|am}}{E_{m|ac}}=1+\frac{2\alpha^{2}\left(e^{2r_{\alpha}}-1\right)}{2\alpha^{2}\beta^{2}\left(e^{r_{\alpha}}-1\right)^{2}+1}.\label{eq19c}
\end{align}
Obviously the ratio is always greater than $1$, so $E_{c|am}>E_{m|ac}$.
This implies that a negative rather than a positive correlation between
steering modes reduces the ability of the modes for steering. It should
be mentioned that the negative correlation between two modes may result
in an entanglement between these modes. Thus, one can conclude that
entanglement between the steering modes leads to a reduction of the
ability of these modes to steer the other mode.

\subsection{Collective tripartite steering}

We now turn to an interesting problem of collective tripartite steering
which occurs when a given mode $i$ is steered by the remaining modes
collectively, $E_{i|jk}<1/2$, and simultaneously is not steered by
each of the modes alone, $E_{i|j}\geq1/2$ and $E_{i|k}\geq1/2$. 

Therefore, to examine the occurrence of collective tripartite steering
we must look at the properties of the bipartite steering parameters
$E_{i|j}$ defined in Eq.~(\ref{bi-steering}). They are readily
evaluated using the solutions for the variances and correlation functions
of the output fields, Eq. (\ref{eq17}). We then obtain analytical
expressions for the bipartite steering parameters. For the case of
steering the cavity mode $a$, we have 
\begin{align}
E_{a|m} & =\Delta(X_{a}+u_{m}P_{m})\Delta(P_{a}+u_{m}X_{m})\nonumber \\
 & =\left(n_{1}\!+\!\frac{1}{2}\right)\!\left[1-\frac{\alpha^{2}(2\bar{n}\!+\!1)\!\left(e^{2r_{\alpha}}\!-\!1\right)}{\Delta^{2}P_{m}}\right],\label{bam}
\end{align}
and 
\begin{align}
E_{a|c} & =\Delta(X_{a}+u_{c}P_{c})\Delta(P_{a}-u_{c}X_{c})\nonumber \\
 & =\left(n_{1}\!+\!\frac{1}{2}\right)\!\left[1+\frac{\alpha^{2}(2\bar{n}\!+\!1)\!\left(e^{2r_{\alpha}}\!-\!1\right)}{\Delta^{2}P_{c}}\right].\label{bac}
\end{align}
For steering of the atomic mode $c$ by cavity mode $a$, we get 
\begin{align}
E_{c|a} & =\Delta(X_{c}+u_{a}P_{a})\Delta(P_{c}-u_{a}X_{a})\nonumber \\
 & =\left(n_{1}\!+\!\frac{1}{2}\right)\!\left[1+\frac{\alpha^{2}\beta^{2}(2\bar{n}+1)\!\left(e^{r_{\alpha}}\!-\!1\right)^{2}}{\Delta^{2}X_{a}}\right],\label{bca}
\end{align}
and by mirror mode $m$ 
\begin{align}
E_{c|m} & =\Delta(X_{c}+u_{m}X_{m})\Delta(P_{c}-u_{m}P_{m})\nonumber \\
 & =\left(n_{1}\!+\!\frac{1}{2}\right)\!\left[1-\frac{\alpha^{2}\beta^{2}(2\bar{n}+1)\left(e^{r_{\alpha}}\!-\!1\right)^{2}}{\Delta^{2}X_{m}}\right].\label{bcm}
\end{align}
Finally, for steering of the mirror mode $m$ by cavity mode $a$,
we find 
\begin{align}
E_{m|a} & =\Delta(X_{m}+u_{a}P_{a})\Delta(P_{m}+u_{a}X_{a})\nonumber \\
 & =\left(n_{0}\!+\!\frac{1}{2}\right)\!\left\{ 1-\left[1-\frac{(2n_{1}+1)(e^{r_{\alpha}}-1)}{(2n_{0}+1)(e^{r_{\alpha}}+1)}\beta^{2}\right]\right.\nonumber \\
 & \left.\times\,\frac{\alpha^{2}(2\bar{n}+1)\left(e^{2r_{\alpha}}\!-\!1\right)}{\Delta^{2}P_{a}}\right\} ,\label{bma}
\end{align}
and by atomic mode $c$ 
\begin{align}
E_{m|c} & =\Delta(X_{m}+u_{c}X_{c})\Delta(P_{m}-u_{c}P_{c})\nonumber \\
 & =\left(n_{0}\!+\!\frac{1}{2}\right)\!\left\{ 1+\left[\frac{(2n_{1}+1)}{(2n_{0}+1)}-\frac{(e^{r_{\alpha}}-1)}{(e^{r_{\alpha}}+1)}\beta^{2}\right]\right.\nonumber \\
 & \left.\times\,\frac{\alpha^{2}(2\bar{n}+1)\left(e^{2r_{\alpha}}\!-\!1\right)}{\Delta^{2}P_{c}}\right\} .\label{bmc}
\end{align}

First, we note that the bipartite parameters (\ref{bam})-(\ref{bmc}),
in the limit $r_{\alpha}\rightarrow\infty$, satisfy the inequality
\begin{equation}
E_{i|j}E_{i|k}=\left(n_{1}+\frac{1}{2}\right)^{2}\geq\frac{1}{4},
\end{equation}
which shows that the bipartite steering properties are in accordance
with the monogamy relation that mode $i$ cannot be simultaneously
steered by modes $j$ and $k$ \cite{mdr113}.

\begin{figure}
\begin{centering}
\includegraphics[width=0.75\columnwidth]{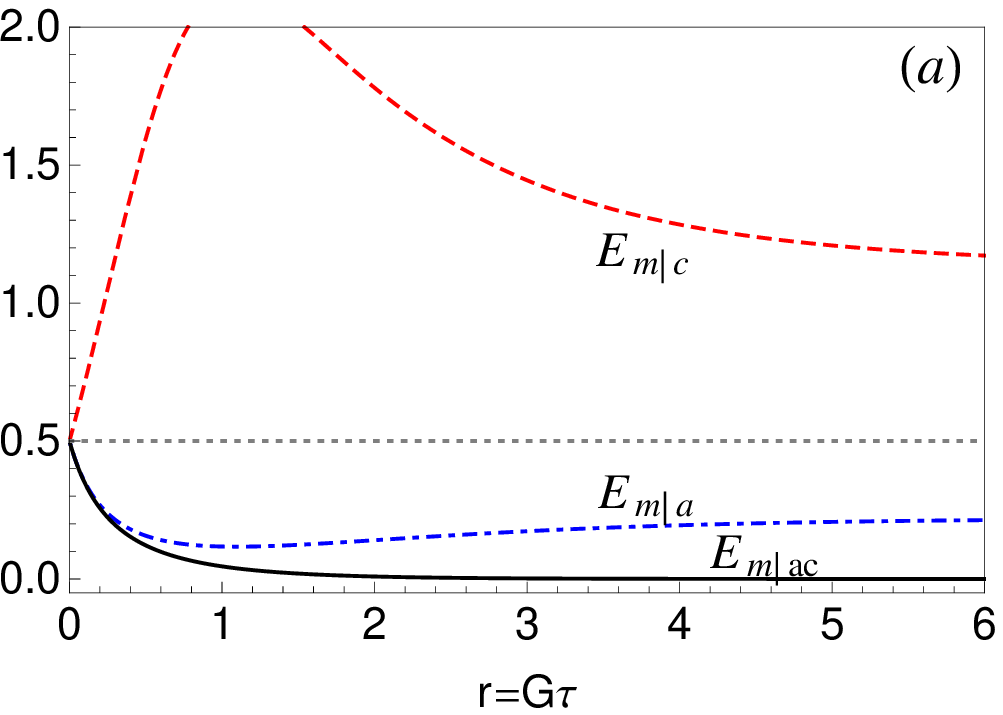} 
\par\end{centering}

\begin{centering}
\includegraphics[width=0.75\columnwidth]{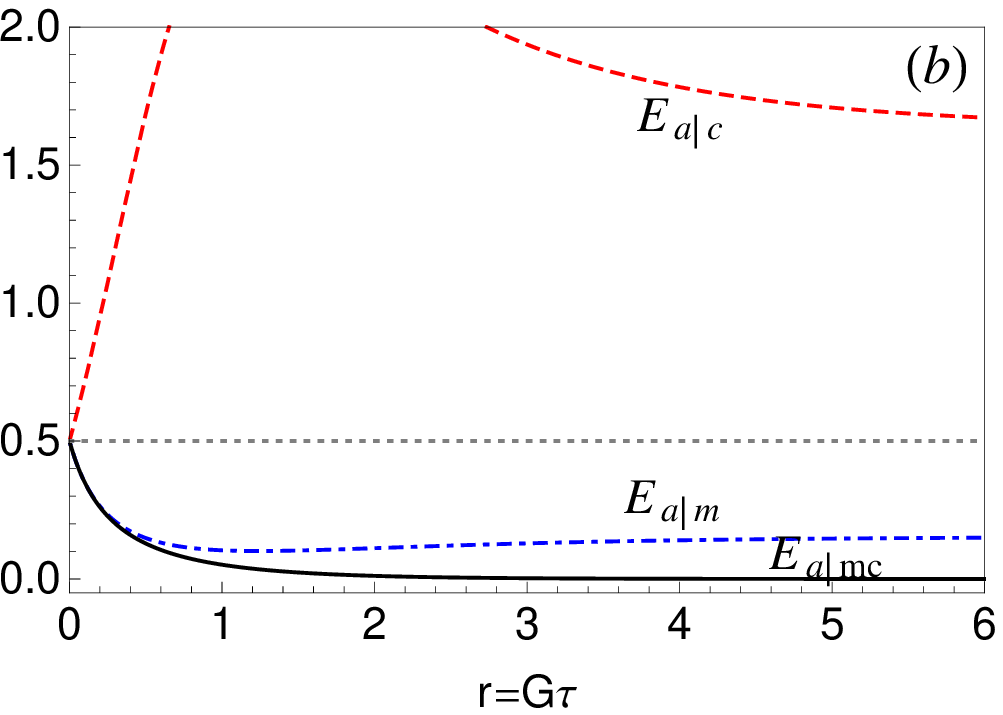} 
\par\end{centering}

\begin{centering}
\includegraphics[width=0.75\columnwidth]{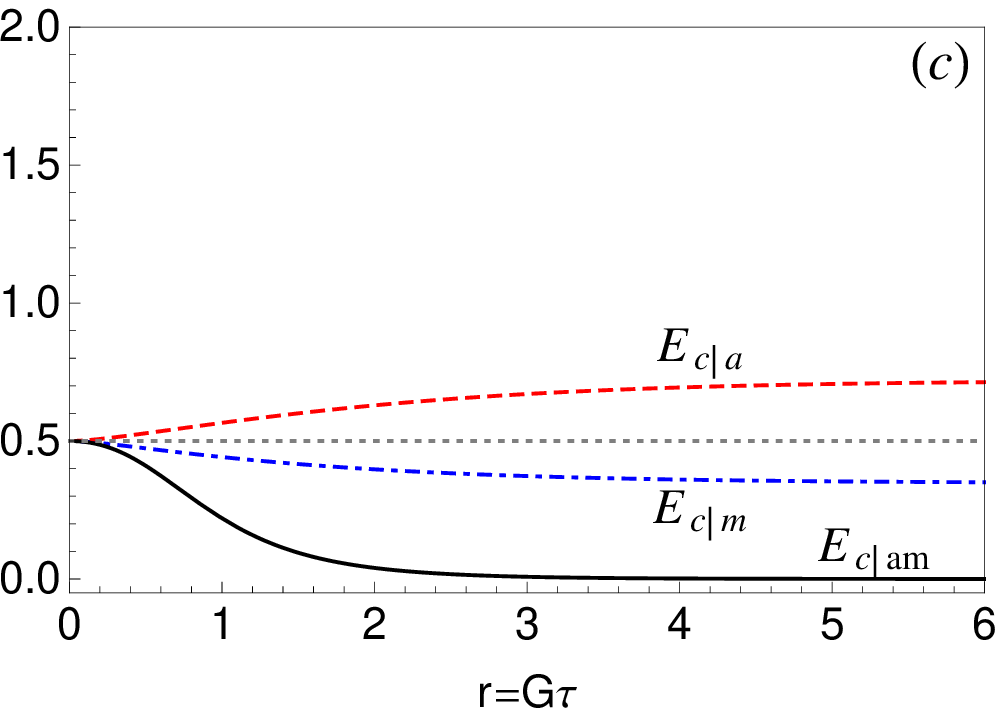} 
\par\end{centering}

\caption{(Color online) Variation of the tripartite and the corresponding bipartite
steering parameters with $r$ for $\alpha=1.2$ and $n_{0}=n_{1}=0$.}

\label{fig:4} 
\end{figure}

From Eqs.~(\ref{bam}-\ref{bmc}), we find that the parameters $E_{a|c}$
and $E_{c|a}$ are always greater than $1/2$. Therefore, we need
only to consider the other four parameters in order to search for
conditions to remove the bipartite steering. It should be noted that
the behavior of the parameters $E_{m|a}$ and $E_{m|c}$, describing
steering properties of the mode $m$, is quite different that either
$E_{m|a}$ or $E_{m|c}$ can be smaller than $1/2$. It depends on
whether $\beta^{2}<(2n_{1}+1)/(2n_{0}+1)$ or $\beta^{2}>1$. For
$\beta^{2}<1$, corresponding to $G_{a}<G/2$, the parameter $E_{m|c}$
is always greater than $1/2$, while $E_{m|a}$ can be reduced below
$1/2$. On the other hand, for $\beta^{2}>1$ this relationship is
reversed and $E_{m|c}$ is the parameter which can be reduced below
$1/2$.

Let us examine in details the dependence of these parameters on $n_{0}$
and $n_{1}$. In the absence of the thermal noise, $n_{0}=n_{1}=0$,
and then $E_{a|m}$ and $E_{c|m}$ are always smaller than $1/2$.
Moreover, depending on whether $\beta<1$ or $\beta>1$, either $E_{m|a}$
or $E_{m|c}$ can be always smaller than $1/2$. Hence, in the absence
of the thermal noise, bipartite steering always occurs.

This is illustrated in Fig.~\ref{fig:4}. We see that the tripartite
steering is accompanied by the bipartite steering over the entire
range of $r$. Thus, the collective steering does not occur. It is
interesting to note that the bipartite steering occurs only between
those modes which are coupled through the parametric interaction.
For example, the modes $a$ and $m$ are directly coupled through
the parametric interaction and it is apparent from the figure that
the modes steer each other. There is no steering between the modes
$a$ and $c$ since the modes are coupled through the beam-splitter
type interaction. The mode $m$ steers the mode $c$ due to the indirect
coupling through the parametric interaction. Note an asymmetry in
the steering between the modes $m$ and $c$ that the mode $c$ is
steered by $m$ ($E_{c|m}<1/2$ shown in Fig. \ref{fig:4}(c)) but
the mode $m$ is not steered by $c$ ($E_{m|c}>1/2$ shown in Fig.
\ref{fig:4}(a)).

\begin{figure}
\begin{centering}
\includegraphics[width=0.75\columnwidth]{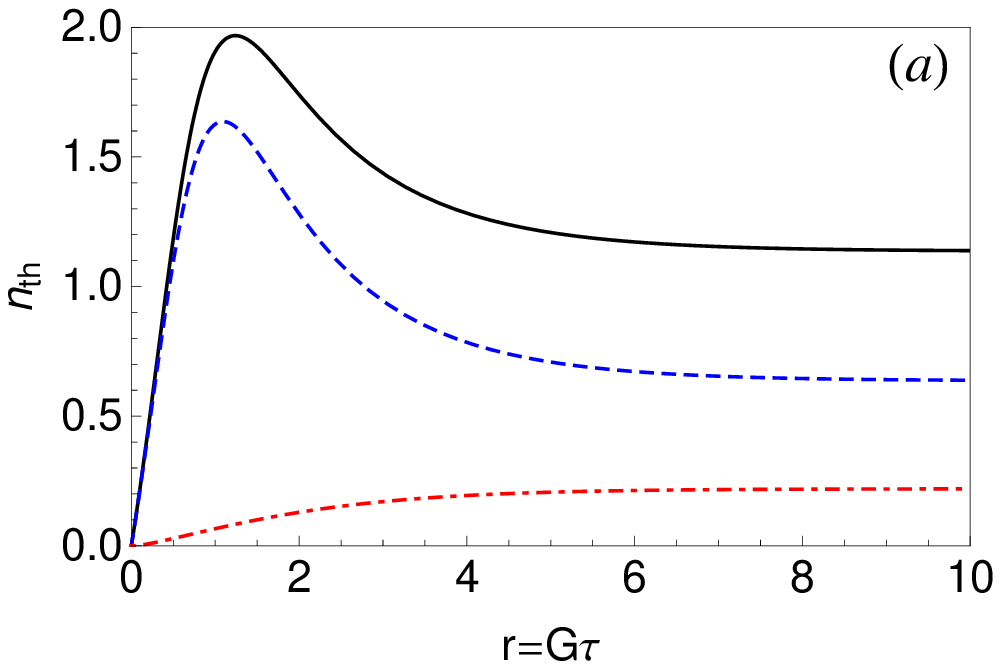} 
\par\end{centering}

\begin{centering}
\includegraphics[width=0.75\columnwidth]{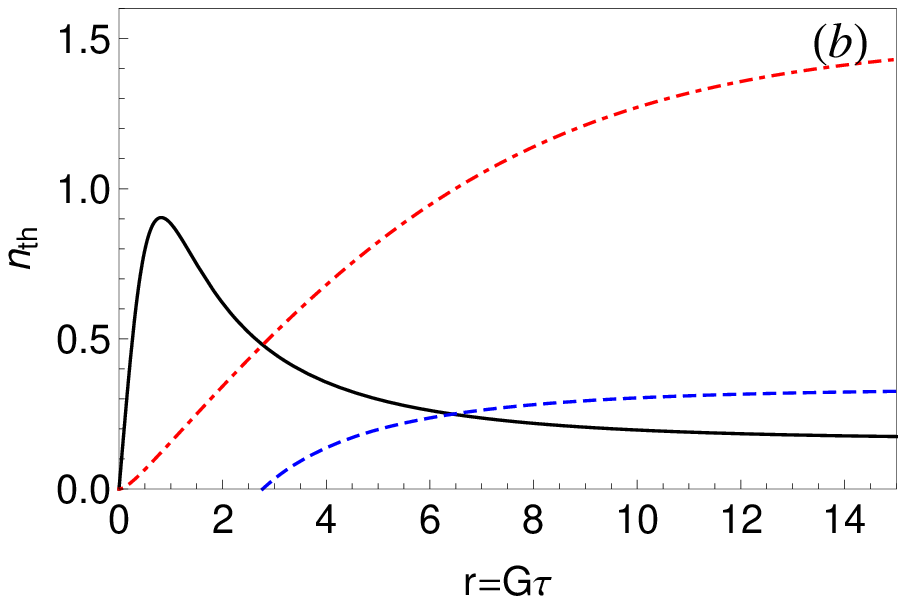} 
\par\end{centering}

\caption{(Color online) Variations of the threshold values $n_{{\rm th}}$
with $r$ for (a) $\alpha=1.2$ and (b) $\alpha=2$. In both frames,
the solid black line is $n_{{\rm th}}$ at which $E_{a|m}=1/2$ and
the red dashed-dotted line shows $n_{{\rm th}}$ at which $E_{c|m}=1/2$.
The blue dashed line in frame (a) represents $n_{{\rm th}}$ at which
$E_{m|a}=1/2$ while in frame (b) it represents $n_{{\rm th}}$ at
which $E_{m|c}=1/2$.}

\label{fig:5} 
\end{figure}

In the presence of the thermal noise, the steering properties of the
modes change dramatically. When all modes are affected by thermal
noise with equal average numbers of photons at each mode, $n_{0}=n_{1}\equiv n\neq0$,
there are minimum (threshold) values for $n$ at which the bipartite
steering parameters become greater than $1/2$. It is easy to see
from Eqs.~(\ref{bam}) and (\ref{bma}), and Eqs. (\ref{bmc}) and
(\ref{bcm}) that the minimum (threshold) values for $n$ are 
\begin{align}
n_{{\rm th}} & =\frac{\alpha^{2}\left(e^{2r_{\alpha}}\!-\!1\right)}{1\!+\!2\alpha^{2}\beta^{2}\!\left(e^{r_{\alpha}}\!-\!1\right)^{2}},\quad{\rm for}\ E_{a|m},\nonumber \\
n_{{\rm th}} & =\frac{\alpha^{2}\!\left(e^{r_{\alpha}}-1\right)\!\left[\alpha^{2}\!+\!(1-\beta^{2})e^{r_{\alpha}}\right]}{1+2\alpha^{2}\beta^{2}\left(e^{r_{\alpha}}-1\right)^{2}},\quad{\rm for}\ E_{m|a},\nonumber \\
n_{{\rm th}} & =\frac{\alpha^{2}\!\left(e^{r_{\alpha}}-1\right)\!\left[(\beta^{2}-1)e^{r_{\alpha}}-\alpha^{2}\right]}{1+2\alpha^{2}\left(e^{2r_{\alpha}}-1\right)},\quad{\rm for}\ E_{m|c},\nonumber \\
n_{{\rm th}} & =\frac{\alpha^{2}\beta^{2}\!\left(e^{r_{\alpha}}-1\right)^{2}}{1+2\alpha^{2}\!\left(e^{2r_{\alpha}}\!-\!1\right)},\quad{\rm for}\ E_{c|m}.\label{eq32}
\end{align}
Note that, in contrast with the corresponding threshold values for
tripartite steering, Eq.~(\ref{eq19u}), the above threshold values
for bipartite steering do not increase exponentially with $r$. In
other words, the threshold values cannot be made arbitrarily large,
they rather saturate at finite values as $r\rightarrow\infty$. This
makes it possible to wipe out the bipartite steering even at small
$n$. This is demonstrated in Fig.~\ref{fig:5} which shows the variation
of the threshold value of $n$ with the squeezing parameter $r$ for
$\alpha=1.2$, corresponding to $\beta<1$, and $\alpha=2$, corresponding
to $\beta>1$.

For $n$ beyond $n=2$, which is above the thresholds defined by Eq.~(\ref{eq32}),
all bipartite steering parameters then go above $1/2$ that the bipartite
steering becomes impossible. Thus, the bipartite steering in the three-mode
optomechanical system can be wiped out by relatively low thermal noise
even in the limit of large squeezing,~$r\rightarrow\infty$.

\begin{figure}
\begin{centering}
\includegraphics[width=0.75\columnwidth]{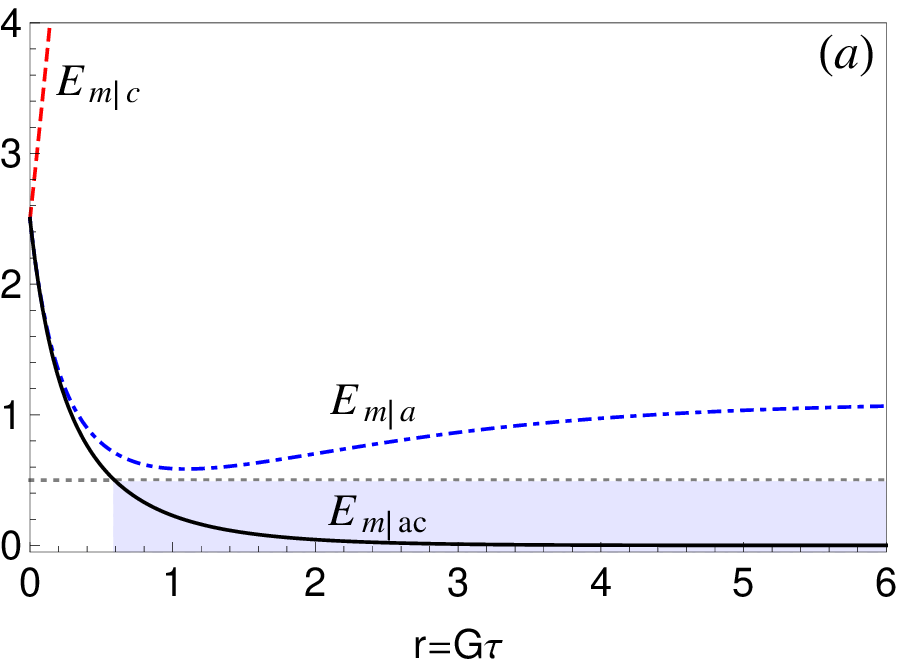} 
\par\end{centering}

\begin{centering}
\includegraphics[width=0.75\columnwidth]{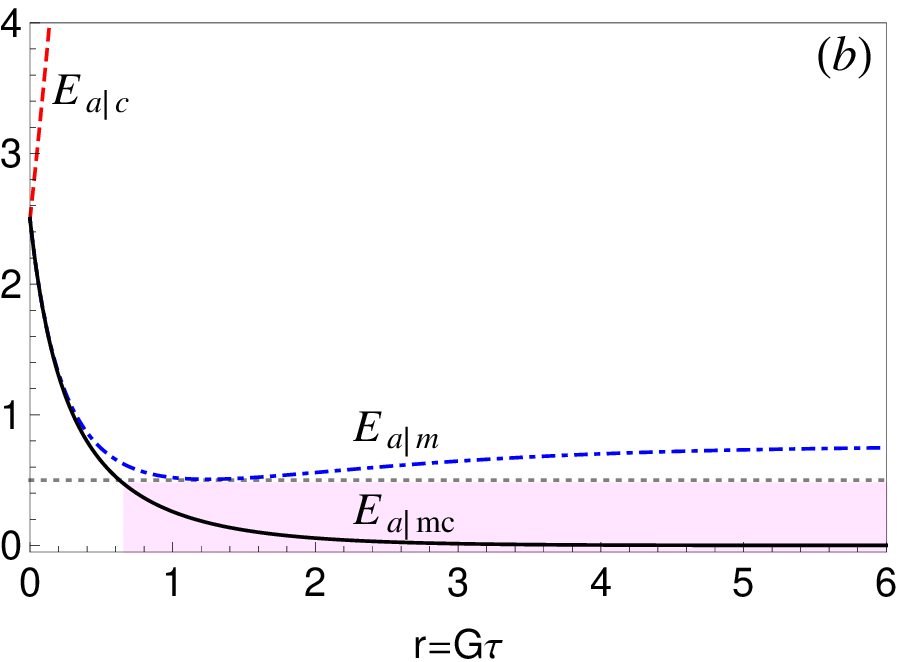} 
\par\end{centering}

\begin{centering}
\includegraphics[width=0.75\columnwidth]{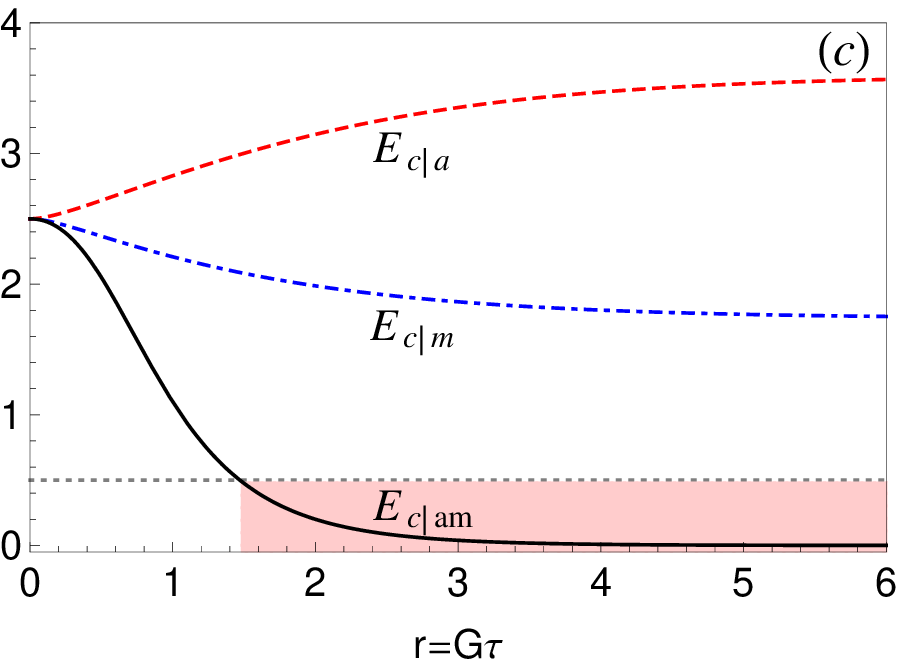} 
\par\end{centering}

\caption{(Color online) Variation of the bipartite and tripartite steering
parameters with $r=G\tau$ for $\alpha=1.2$ and the thermal noise
present at three modes with equal average number of photons, $n_{0}=n_{1}=2$.
The shaded region marks the range of the squeezing parameter $r$
over which a given mode can be steered collectively by the remaining
two modes. \label{fig:6}}
\end{figure}

Although the effect of steering of mode $i$ separately by the modes
$j$ and $k$ disappears when $n$ is larger than the threshold value,
it must not be thought that all steering effects then disappear. The
steering property may still be there, but it could be manifested in
collective rather than individual steering behavior of the modes $j$
and $k$, as shown in Fig. \ref{fig:6}. The shaded areas indicate
the range of $r$ over which the collective steering of a given mode
by the collection of the remaining modes. We see that the conditions
for collective steering of the modes are satisfied everywhere except
for very small squeezing. It is clear from the figure that unlike
the tripartite steering, the bipartite steering can be removed by
the thermal noise.

\begin{figure}
\begin{centering}
\includegraphics[width=0.75\columnwidth]{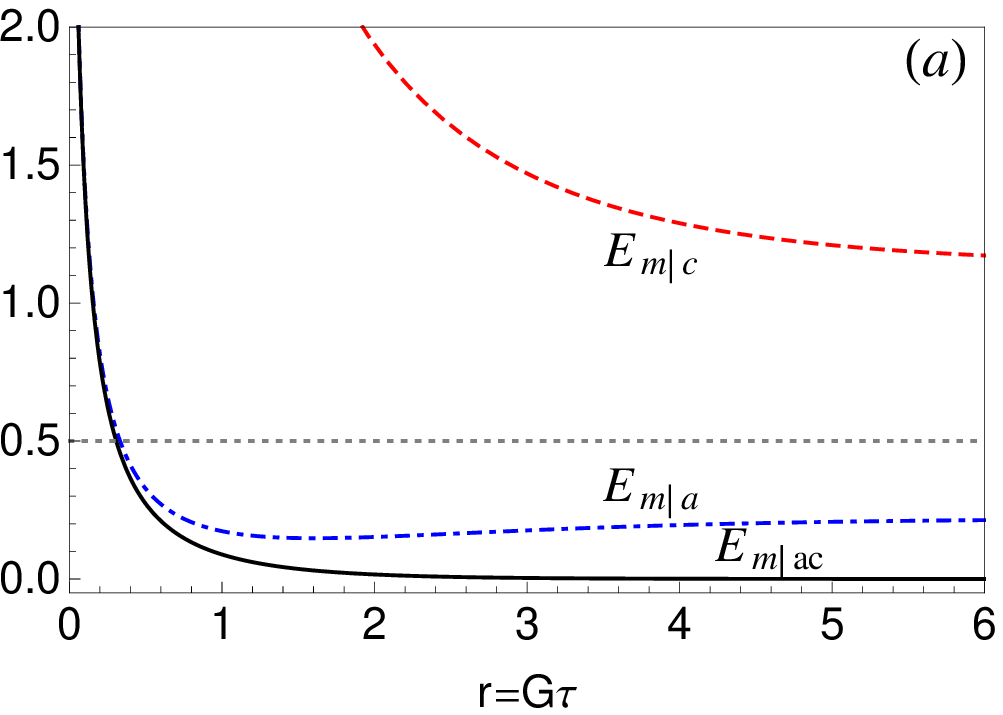} 
\par\end{centering}

\begin{centering}
\includegraphics[width=0.75\columnwidth]{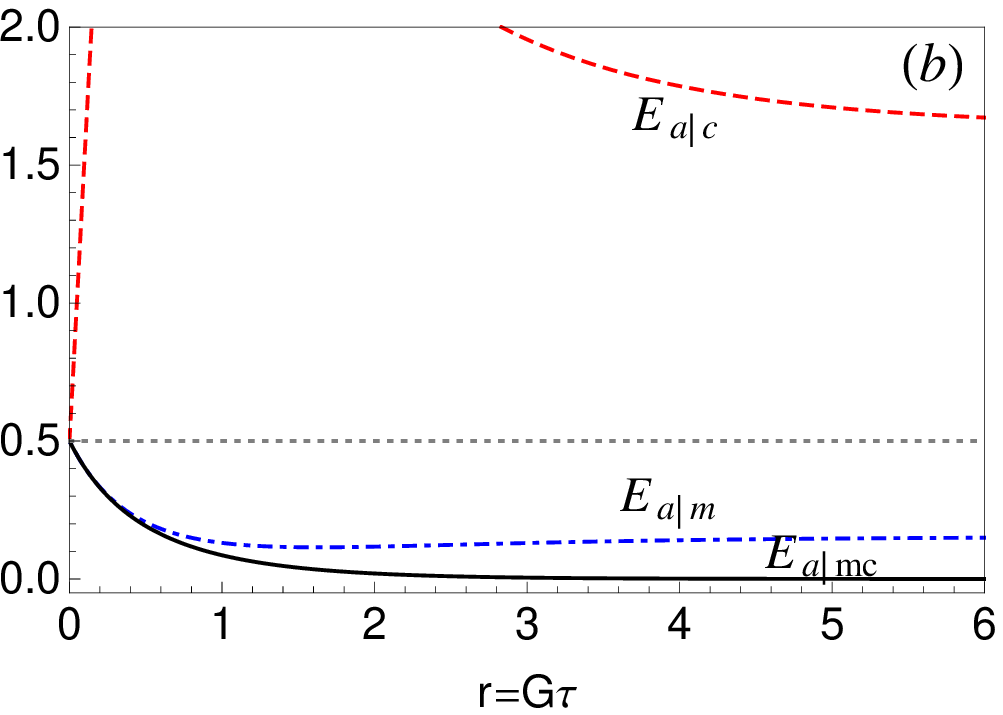} 
\par\end{centering}

\begin{centering}
\includegraphics[width=0.75\columnwidth]{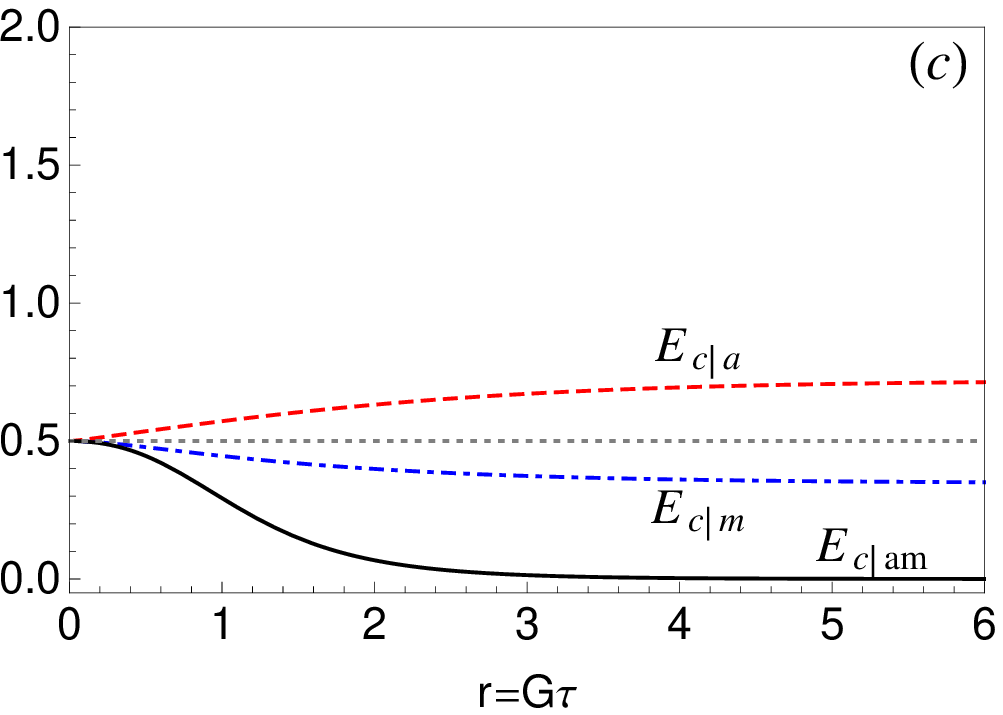} 
\par\end{centering}

\caption{(Color online) Variation of the bipartite and tripartite steering
parameters with $r=G\tau$ for $\alpha=1.2$ and the thermal noise
present only at the mirror mode, $n_{0}=4$ and $n_{1}=0$.\textcolor{red}{{}
\label{fig:7}}}
\end{figure}

One can notice from Fig.~\ref{fig:6} that in the case of thermal
noise affecting all three modes $(n_{0}=n_{1})$, the tripartite and
collective tripartite steering occur over the same range of $r$.
The situation differs if initially only one or two modes of the system
were in the thermal state, either $n_{0}\neq0$, $n_{1}=0$ or $n_{0}=0$,
$n_{1}\neq0$. A close look at the parameters (\ref{bam})-(\ref{bmc})
reveals that similar to the tripartite steering, the bipartite steering
parameters are mostly affected by the thermal noise present initially
in the steered mode. Namely, the steering of the mode $m$ is limited
by the factor $(n_{0}+1/2)$, whereas the steering of the modes $a$
and $c$ is limited by the factor $(n_{1}+1/2)$. This clearly shows
that the bipartite steering can be wiped out only by the thermal noise
present in the steered mode. Equations~(\ref{bam})-(\ref{bmc})
also show that the thermal noise at the steering mode has a marginal
effect on the bipartite steering. It is particularly well seen in
the case of $n_{0}\gg1$, in which the steering parameters $E_{a|m}$,
Eq.~(\ref{bam}), and $E_{c|m}$, Eq.~(\ref{bcm}), can be simplified~to
\begin{align}
E_{a|m}=\left(n_{1}\!+\!\frac{1}{2}\right)\!\left[1-\frac{\alpha^{2}\!\left(e^{2r_{\alpha}}\!-\!1\right)}{\left(\alpha^{2}e^{r_{\alpha}}-\beta^{2}\right)^{2}}\right],\label{bam1}
\end{align}
and 
\begin{align}
E_{c|m}=\left(n_{1}\!+\!\frac{1}{2}\right)\!\left[1-\frac{\alpha^{2}\beta^{2}\left(e^{r_{\alpha}}\!-\!1\right)^{2}}{\left(\alpha^{2}e^{r_{\alpha}}-\beta^{2}\right)^{2}}\right].\label{bcm1}
\end{align}
Evidently, the parameters are independent of $n_{0}$, and could be
larger than $1/2$ only if $n_{1}\neq0$.

\begin{figure}
\begin{centering}
\includegraphics[width=0.75\columnwidth]{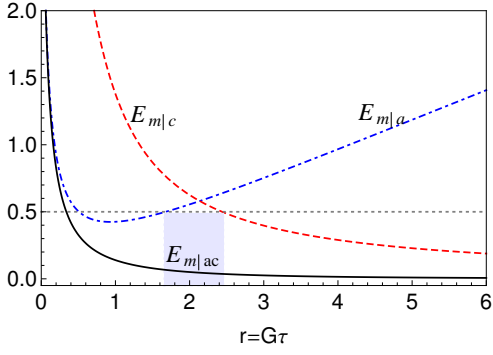} 
\par\end{centering}

\caption{(Color online) Variation of the bipartite and tripartite steering
parameters with $r=G\tau$ for $\alpha=4$ and the thermal noise initially
present only in the mirror mode, $n_{0}=4$ and $n_{1}=0$. The shaded
region marks the range of the squeezing parameter $r$ over which
the mode $m$ can be steered collectively by the modes $a$ and $c$.\label{fig:8}}
\end{figure}

The variation of the bipartite and tripartite steering parameters
with $r$ for the initial thermal state of the mode $m$, $n_{0}=4$,
and vacuum state of the other modes, $n_{1}=0$ is illustrated in
Fig.~\ref{fig:7}. We see that in Figs.~\ref{fig:7}(b) and \ref{fig:7}(c)
where the mode $m$ appears as a steering mode, the tripartite steering
then occurs at the entire range of $r$. Thus, thermal noise present
solely at one of the steering modes cannot wipe out all the bipartite
steering in the system. As a result, there is no collective steering
possible. No collective steering is seen even in the case where the
mode $m$ is steered by the other modes, Fig.~\ref{fig:7}(a). In
fact, it is possible for $E_{m|a}$ to be greater than $1/2$ at some
values of $r$, so that the collective steering of the mode $m$ could
occur at those values of $r$. This, however, requires larger values
of either $n_{0}$ or $\alpha$. Figure~\ref{fig:8} shows the variation
of the bipartite steering parameters, $E_{m|c},E_{m|a}$, and the
tripartite steering parameter $E_{m|ac}$ with $r$ for a larger value
of $\alpha=4$. We see that the bipartite steering can be removed
at a very restricted range of $r$ with no significant changes to
the tripartite steering. At that range of $r$, the tripartite steering
corresponds to the collective steering. 

\begin{figure}
\begin{centering}
\includegraphics[width=0.75\columnwidth]{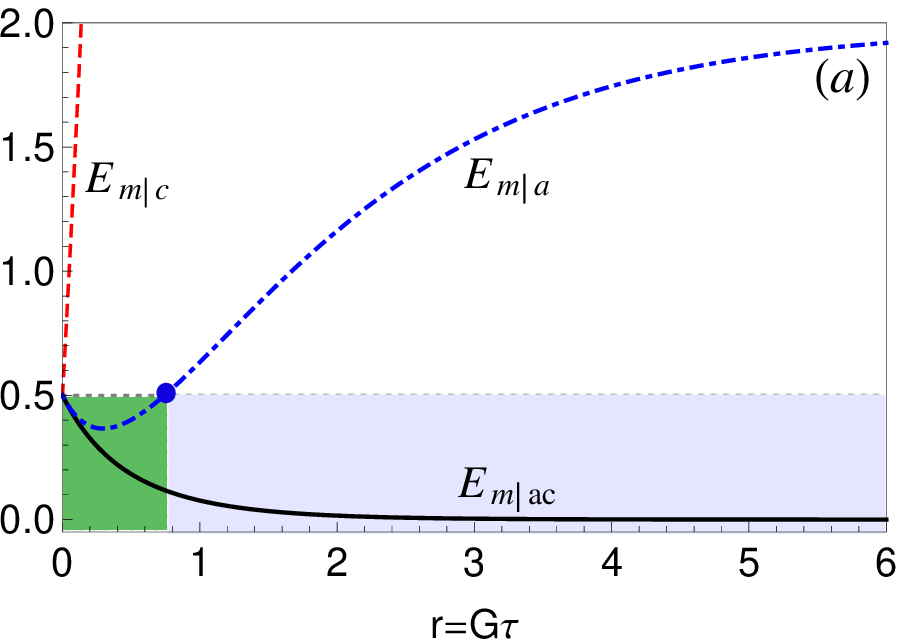} 
\par\end{centering}

\begin{centering}
\includegraphics[width=0.75\columnwidth]{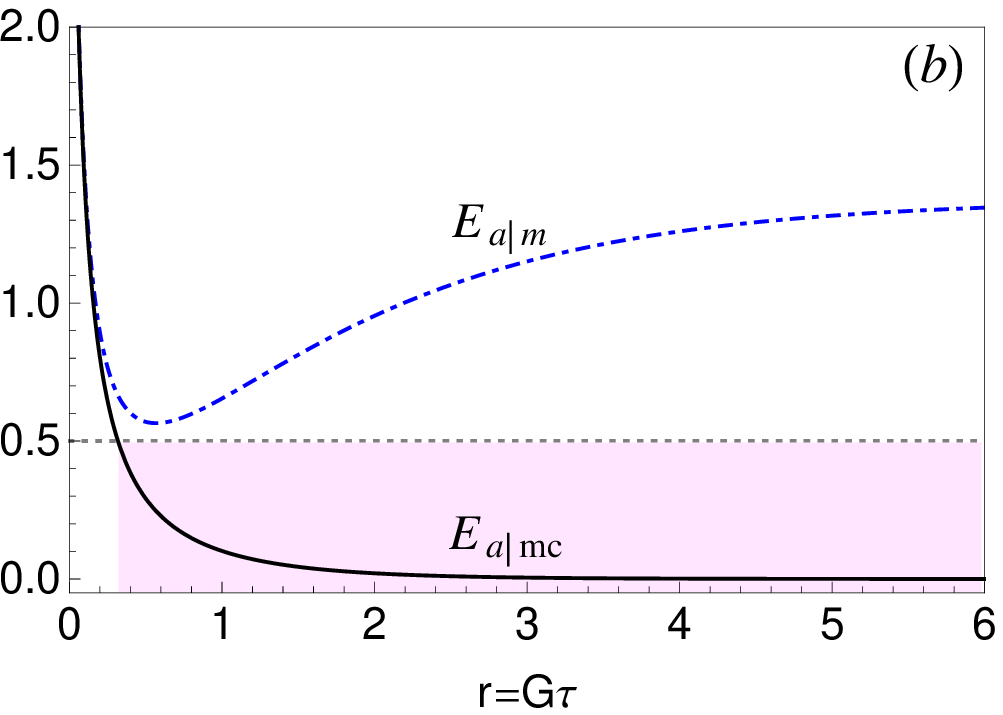} 
\par\end{centering}

\begin{centering}
\includegraphics[width=0.75\columnwidth]{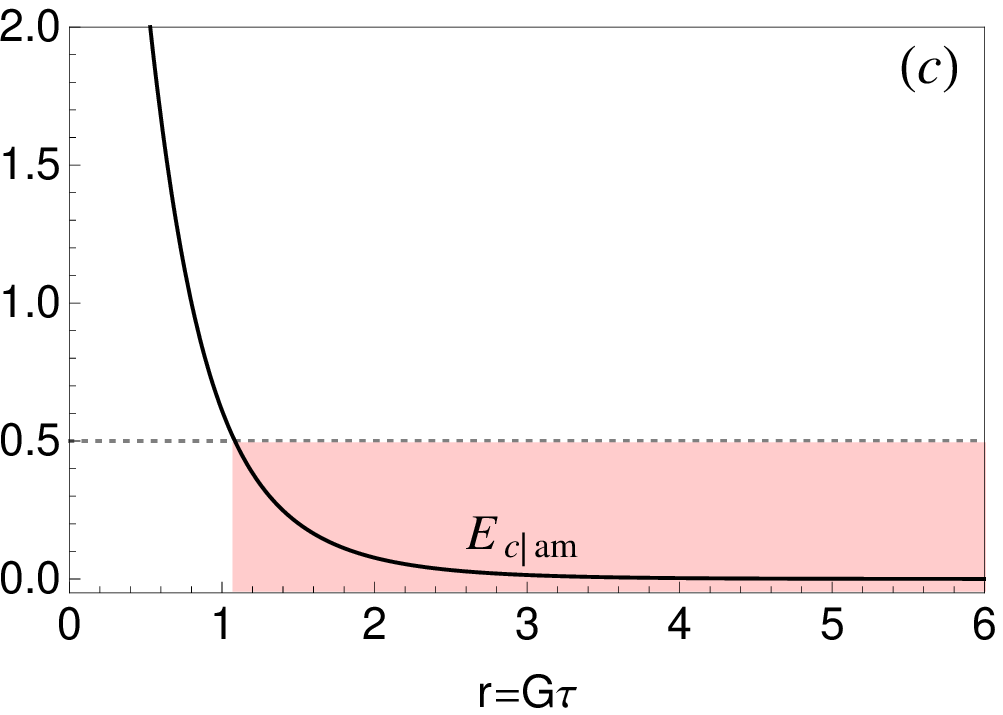} 
\par\end{centering}

\caption{(Color online) Variation of the bipartite and tripartite steering
parameters with $r=G\tau$ for $\alpha=1.2$ and the thermal noise
present at two modes, the cavity and atomic modes, $n_{0}=0$ and
$n_{1}=4$. The shaded region marks the range of the squeezing parameter
$r$ over which a given mode can be steered collectively by the remaining
two modes. The narrower green region (before the blue point) in (a)
marks the range of $r$ where the tripartite steering is present but
the collective steering can never occur. \label{fig:9}}
\end{figure}

Figure~\ref{fig:9} shows the variation of the bipartite and tripartite
steering parameters with $r$ for the case when two modes of the system,
$a$ and $c$ are initially in a thermal state, i.e, $n_{0}=0$ and
$n_{1}=4$. We see from Fig.~\ref{fig:9}(a) that if the steering
mode is initially in a thermal state, the tripartite steering takes
place over the entire range of $r$ and the collective steering is
seen to occurs but it is restricted to a finite range of $r$. There
is a range of $r$ at which the collective steering is impossible.
Thus, the thermal noise initially present in the mode $a$ cannot
remove the bipartite steering of the mode $m$. Only in the limit
of $n_{1}\rightarrow\infty$ the parameter $E_{m|a}$ goes to $1/2$.
In other words, thermal noise present only in the steering mode can
never completely remove all the bipartite steering. In this case the
tripartite steering cannot be always regarded as the collective steering.
In the opposite situation, shown in Fig.~\ref{fig:9}(b) where initially
the steered mode $a$ and steering mode $c$, and in Fig.~\ref{fig:9}(c)
where initially the steered mode $c$ and steering mode $a$ were
in the thermal state, both tripartite and collective steering occur
in the same range of $r$. Thus, if the steered and one of the steering
modes were initially in the thermal state, the entire tripartite steering
could coincide with the collective steering with appropriate values
of $\alpha$ and $n_{1}$.

We may summarize that the results presented in Figs.~\ref{fig:6}-\ref{fig:9}
clearly show that tripartite steering always occurs in the system
regardless of the initial state of the modes, but collective tripartite
steering can occur only if the modes are initially in a thermal state.
The questions whether the tripartite steering coincides with the collective
steering depends strongly on the redistribution of the thermal noise
between the modes.

In closing this section, we briefly comment on the role of collective
tripartite steering in security of hybrid quantum networks as a resource
for quantum secret~sharing \cite{hb99,ag12,lc14}. Suppose Alice
wishes to transmit a secret message to two parties, Bob and Charlie.
Before transmitting the message, Alice may send a quantum encryption
key separately to Bob and Charlie or she can distribute the key among
them, so in the later case Bob and Charlie must collaborate to decipher
the message. The important feature is that steering ability, unlike
ordinary entanglement, is constrained by the violation of inferred
Heisenberg relation, $\Delta_{inf,j}X_{i}\Delta_{inf,j}P_{i}\geq1/2$,
which cannot be performed by classical means. When receiving the message
from Alice (the result of Alice's measurement), Bob and Charlie have
to deduce the value of the amplitude of Alice's system by demonstrating
the violation of inferred Heisenberg uncertainty relation by collective
measurements on their systems.

\section{Conclusions}

\label{Sec5}

We have studied the steering properties of the bosonic modes of a
hybrid pulsed cavity optomechanical system composed of a single-mode
cavity with a movable fully reflective mirror and containing an ensemble
of two-level atoms. The cavity mode was driven by light pulses and
the variances and correlation functions of the amplitudes of the output
modes were evaluated. The treatment was restricted to the bad cavity
limit under which the adiabatic approximation was made of a slowly
varying amplitude of the cavity mode. The laser pulses were assumed
to be strong and blue detuned to the cavity and the atomic resonance
frequencies. The solutions were then used to obtain analytic expressions
for the steering parameters. We were particularly interested in the
dependence of the bipartite and tripartite steering on the initial
state of the modes. We have found that the initial thermal noise presented
in the modes is more effective in destroying the bipartite rather
than the tripartite steering. In fact, the tripartite steering can
persist even for a large thermal noise. When the bipartite steering
is destroyed then the existing tripartite steering can be regarded
as collective tripartite steering. A detailed analysis has shown that
the occurrence of the collective tripartite steering is highly sensitive
to number of modes being initially in thermal states and to whether
the noise affected mode appears as a steered or steering mode. In
the case where initially only a steered mode is in a thermal state,
the bipartite steering could be completely destroyed and then the
existing tripartite steering corresponds to the collective steering.
On the other hand, when initially only the steering modes were in
thermal states, the collective steering of the remaining mode may
occur but it is present in a very restricted range of the interaction
time. In particular, the collective steering is absent when only one
of the steering modes was initially in a thermal state. When initially
both steering modes were in a thermal state, a collective steering
may occur in a less restricted range of the interaction time. When
all modes are initially in thermal states, the bipartite steering
can be destroyed completely leaving the collective steering as the
only steering present in the system. We note that the collective steering
has potential application for quantum secret sharing in a hybrid quantum
network. 

\acknowledgments We thank P. Meystre, M. D. Reid, and P. Drummond
for useful discussions. We acknowledge support from the National Natural
Science Foundation of China under Grant Nos. 11274025 and 11121091.

\appendix
%dummy comment inserted by tex2lyx to ensure that this paragraph is not empty
%dummy comment inserted by tex2lyx to ensure that this paragraph is not empty
%dummy comment inserted by tex2lyx to ensure that this paragraph is not empty

\section{}

In this appendix we give the expressions for the bipartite and tripartite
steering parameters in terms of the variances and correlation functions,
and optimal weight factors that minimize the variances involved in
the steering parameters.

\subsection{Bipartite steering parameters and optimal gain factors}

The parameter $E_{m|a}$ determining bipartite steering of the mirror
mode $m$ by the cavity mode $a$ is determined by
\begin{align}
E_{m|a} & =\Delta\left[X_{m}+u_{a}P_{a}\right]\Delta\left[P_{m}+u_{a}X_{a}\right]\nonumber \\
 & =\Delta^{2}X_{m}+u_{a}^{2}\Delta^{2}P_{a}+2u_{a}\left\langle X_{m},P_{a}\right\rangle ,
\end{align}
where$\left\langle X_{m},P_{a}\right\rangle =\frac{1}{2}\left\langle X_{m}P_{a}+P_{a}X_{m}\right\rangle -\left\langle X_{m}\right\rangle \left\langle P_{a}\right\rangle $.
This can be minimized with the optimal weight factor 
\begin{equation}
u_{a}=-\frac{\left\langle X_{m},P_{a}\right\rangle }{\Delta^{2}P_{a}}.
\end{equation}

The steering of the mirror mode $m$ by the atomic mode $c$ is given
by 
\begin{align}
E_{m|c} & =\Delta\left[X_{m}+u_{c}X_{c}\right]\Delta\left[P_{m}-u_{c}P_{c}\right]\nonumber \\
 & =\Delta^{2}X_{m}+u_{c}^{2}\Delta^{2}X_{c}+2u_{c}\left\langle X_{m},X_{c}\right\rangle ,
\end{align}
which can be minimized by the optimal weight factor 
\begin{equation}
u_{c}=-\frac{\left\langle X_{m},X_{c}\right\rangle }{\Delta^{2}X_{c}}.
\end{equation}

The steering of the cavity mode $a$ by the mirror mode $m$ is given
by 
\begin{align}
E_{a|m} & =\Delta\left[X_{a}+u_{m}P_{m}\right]\Delta\left[P_{a}+u_{m}X_{m}\right]\nonumber \\
 & =\Delta^{2}X_{a}+u_{m}^{2}\Delta^{2}P_{m}+2u_{m}\left\langle X_{a},P_{m}\right\rangle ,
\end{align}
which can be minimized by the optimal weight factor 
\begin{equation}
u_{m}=-\frac{\left\langle X_{a},P_{m}\right\rangle }{\Delta^{2}P_{m}}
\end{equation}

The steering of the cavity mode $a$ by the atomic mode $c$ is given
by 
\begin{align}
E_{a|c} & =\Delta\left[X_{a}+u_{c}P_{c}\right]\Delta\left[P_{a}-u_{c}X_{c}\right]\nonumber \\
 & =\Delta^{2}X_{a}+u_{c}^{2}\Delta^{2}P_{c}+2u_{c}\left\langle X_{a},P_{c}\right\rangle ,
\end{align}
which can be minimized by the optimal weight factor 
\begin{equation}
u_{c}=-\frac{\left\langle X_{a},P_{c}\right\rangle }{\Delta^{2}P_{c}}.
\end{equation}

The steering of atomic mode $c$ by the cavity mode $a$ is given
by 
\begin{align}
E_{c|a} & =\Delta\left[X_{c}+u_{a}P_{a}\right]\Delta\left[P_{c}-u_{a}X_{a}\right]\nonumber \\
 & =\Delta^{2}X_{c}+u_{a}^{2}\Delta^{2}P_{a}+2u_{a}\left\langle X_{c},P_{a}\right\rangle ,
\end{align}
which can be minimized by the optimal weight factor 
\begin{equation}
u_{a}=-\frac{\left\langle X_{c},P_{a}\right\rangle }{\Delta^{2}P_{a}}
\end{equation}

The steering of atomic mode $c$ by the mirror mode $m$ is given
by 
\begin{align}
E_{c|m} & =\Delta\left[X_{c}+u_{m}X_{m}\right]\Delta\left[P_{c}-u_{m}P_{m}\right]\nonumber \\
 & =\Delta^{2}X_{c}+u_{m}^{2}\Delta^{2}X_{m}+2u_{m}\left\langle X_{c},X_{m}\right\rangle ,
\end{align}
which can be minimized by the optimal weight factor 
\begin{equation}
u_{m}=-\frac{\left\langle X_{c},X_{m}\right\rangle }{\Delta^{2}X_{m}}.
\end{equation}

\subsection{Tripartite steering parameters and optimal gain factors}

The steering of the mirror mode $m$ by the group $\{ac\}$ is determined
by the parameter 
\begin{align}
E_{m|ac} & =\Delta\!\left[X_{m}\!+\!\left(u_{a}P_{a}\!+\! u_{c}X_{c}\right)\right]\Delta\!\left[P_{m}+\left(u_{a}X_{a}\!-\! u_{c}P_{c}\right)\right]\nonumber \\
 & =\Delta^{2}X_{m}+u_{a}^{2}\Delta^{2}P_{a}+u_{c}^{2}\Delta^{2}X_{c}+2u_{a}\left\langle X_{m},P_{a}\right\rangle \nonumber \\
 & \ \ \ \ +2u_{c}\left\langle X_{m},X_{c}\right\rangle +2u_{a}u_{c}\left\langle P_{a},X_{c}\right\rangle ,
\end{align}
which can be minimized by the optimal weight factors 
\begin{align}
u_{a} & =\frac{\Delta^{2}X_{c}\left\langle X_{m},P_{a}\right\rangle -\left\langle X_{m},X_{c}\right\rangle \left\langle P_{a},X_{c}\right\rangle }{\left\langle P_{a},X_{c}\right\rangle ^{2}-\Delta^{2}P_{a}\Delta^{2}X_{c}},\nonumber \\
u_{c} & =\frac{\Delta^{2}P_{a}\left\langle X_{m},X_{c}\right\rangle -\left\langle X_{m},P_{a}\right\rangle \left\langle P_{a},X_{c}\right\rangle }{\left\langle P_{a},X_{c}\right\rangle ^{2}-\Delta^{2}P_{a}\Delta^{2}X_{c}}.
\end{align}

The parameter $E_{a|mc}$ determining tripartite steering of the cavity
mode $a$ by the group $\{mc\}$ is given by 
\begin{align}
E_{a|mc} & =\Delta\!\left[X_{a}\!+\!\left(u_{m}P_{m}\!+\! u_{c}P_{c}\right)\right]\Delta\!\left[P_{a}\!+\!\left(u_{m}X_{m}\!-\! u_{c}X_{c}\right)\right]\nonumber \\
 & =\Delta^{2}X_{a}+u_{m}^{2}\Delta^{2}P_{m}+u_{c}^{2}\Delta^{2}P_{c}+2u_{m}\left\langle X_{a},P_{m}\right\rangle \nonumber \\
 & \ \ \ \ +2u_{c}\left\langle X_{a},P_{c}\right\rangle +2u_{m}u_{c}\left\langle P_{m},P_{c}\right\rangle ,
\end{align}
which can be minimized by the optimal weight factors 
\begin{align}
u_{m} & =\frac{\Delta^{2}P_{c}\left\langle X_{a},P_{m}\right\rangle -\left\langle X_{a},P_{c}\right\rangle \left\langle P_{m},P_{c}\right\rangle }{\left\langle P_{m},P_{c}\right\rangle ^{2}-\Delta^{2}P_{m}\Delta^{2}X_{c}},\nonumber \\
u_{c} & =\frac{\Delta^{2}P_{m}\left\langle X_{a},P_{c}\right\rangle -\left\langle X_{a},P_{m}\right\rangle \left\langle P_{m},P_{c}\right\rangle }{\left\langle P_{m},P_{c}\right\rangle ^{2}-\Delta^{2}P_{m}\Delta^{2}X_{c}}.
\end{align}

Finally, the parameter $E_{c|am}$ determining tripartite steering
of the atomic mode $c$ by the group $\{am\}$ is given~by 
\begin{align}
E_{c|am} & =\Delta\!\left[X_{c}\!+\!\left(u_{a}P_{a}\!+\! u_{m}X_{m}\right)\right]\Delta\!\left[P_{c}\!-\!\left(u_{a}X_{a}\!+\! u_{m}P_{m}\right)\right]\nonumber \\
 & =\Delta^{2}X_{c}+u_{a}^{2}\Delta^{2}P_{a}+u_{m}^{2}\Delta^{2}X_{m}+2u_{a}\left\langle X_{c},P_{a}\right\rangle \nonumber \\
 & \ \ \ \ +2u_{m}\left\langle X_{c},X_{m}\right\rangle +2u_{a}u_{m}\left\langle P_{a},X_{m}\right\rangle ,
\end{align}
which can be minimized by the optimal weight factors 
\begin{align}
u_{a} & =\frac{\Delta^{2}X_{m}\left\langle X_{c},P_{a}\right\rangle -\left\langle X_{c},X_{m}\right\rangle \left\langle P_{a},X_{m}\right\rangle }{\left\langle P_{a},X_{m}\right\rangle ^{2}-\Delta^{2}P_{a}\Delta^{2}X_{m}},\nonumber \\
u_{m} & =\frac{\Delta^{2}P_{a}\left\langle X_{c},X_{m}\right\rangle -\left\langle X_{c},P_{a}\right\rangle \left\langle P_{a},X_{m}\right\rangle }{\left\langle P_{a},X_{m}\right\rangle ^{2}-\Delta^{2}P_{a}\Delta^{2}X_{m}}.
\end{align}


\begin{thebibliography}{10}
\bibitem{werner} R. F. Werner, Phys. Rev. A \textbf{40}, 4277 (1989).

\bibitem{hw}H. M. Wiseman, S. J. Jones, and A. C. Doherty, Phys.
Rev. Lett. \textbf{98}, 140402 (2007).

\bibitem{jones}S. Jones, H. M. Wiseman, and A. C. Doherty, Phys.
Rev. A \textbf{76}, 052116 (2007).

\bibitem{eprappligrangcopy}C. Branciard, E. G. Cavalcanti, S. P.
Walborn, V. Scarani, and H. M. Wiseman, Phys. Rev. A \textbf{85},
010301(R) (2012).

\bibitem{mdr13} M. D. Reid, Phys. Rev. A \textbf{88}, 062338 (2013).

\bibitem{ec13} D. A. Evans, E. G. Cavalcanti, and H. M. Wiseman,
Phys. Rev. A \textbf{88}, 022106 (2013).

\bibitem{mko13} M. K. Olsen, Phys. Rev. A \textbf{88}, 051802(R)
(2013).

\bibitem{qmNJP}Q. Y. He and M. D. Reid, New J. Phys. \textbf{15},
063027 (2013). 

\bibitem{cp14} Priyanka Chowdhury, Tanumoy Pramanik, A. S. Majumdar, and G. S. Agarwal, Phys. Rev. A \textbf{89}, 012104 (2014).

\bibitem{bv14} J. Bowles, T. Vertesi, M. T. Quintino, and N. Brunner,
Phys. Rev. Lett. \textbf{112}, 200402 (2014).

\bibitem{sn14} P. Skrzypczyk, M. Navascues, and D. Cavalcanti, Phys.
Rev. Lett. \textbf{112}, 180404 (2014).

\bibitem{cl14} Y. N. Chen, C. M. Li, N. Lambert, S. L. Chen, Y. Ota,
G. Y. Chen, and F. Nori, Phys. Rev. A \textbf{89}, 032112 (2014),

\bibitem{sb14} N. Stevens and P. Busch, Phys. Rev. A \textbf{89},
022123 (2014).

\bibitem{einstein}A. Einstein, B. Podolsky, and N. Rosen, Phys. Rev.
\textbf{47}, 777 (1935).

\bibitem{epr1989} M. D. Reid, Phys. Rev. A \textbf{40}, 913 (1989).

\bibitem{rmp}M. D. Reid \textit{et al.}, Rev. Mod. Phys. \textbf{81},
1727 (2009).

\bibitem{Eric2009}E. G. Cavalcanti, S. J. Jones, H. M. Wiseman, and
M. D. Reid, Phys. Rev. A \textbf{80}, 032112 (2009).

\bibitem{oneway steering} V. Handchen {\it et al.}, Nature Photonics \textbf{6},
598 (2012).

\bibitem{genuine steering} Q. Y. He and M. D. Reid, Phys. Rev. Lett.
\textbf{111}, 250403 (2013).

\bibitem{opto-ent}S. Mancini, V. Giovannetti, D. Vitali, and P. Tombesi,
Phys. Rev. Lett. \textbf{88}, 120401 (2002).

\bibitem{ag10} M. Aspelmeyer, S. Groblacher, K. Hammerer, and N.
Kiesel, J. Opt. Soc. Am. B \textbf{27}, A189 (2010).

\bibitem{peter}S. Kiesewetter, Q. Y. He, P. D. Drummond, and M. D.
Reid, arXiv:1312.6474.

\bibitem{HeReid2013pulseent} Q. Y. He and M. D. Reid, Phys. Rev.
A \textbf{88}, 052121 (2013).

\bibitem{Hofer}S. G. Hofer, W. Wieczorek, M. Aspelmeyer, and K. Hammerer,
Phys. Rev. A \textbf{84}, 052327 (2011).

\bibitem{cooling} A. D. O'Connell \textit{et al.}, Nature \textbf{464},
697 (2010); J. Chan \textit{et al.}, Nature \textbf{478}, 89 (2011);
S. Groblacher \textit{et al.}, Nature Phys. \textbf{5}, 485 (2009);
J. D. Teufel \textit{et al.}, Nature \textbf{475}, 359 (2011); \textbf{471},
204 (2011).

\bibitem{zhang2003}J. Zhang, K. Peng, and S. L. Braunstein, Phys.
Rev. A \textbf{68}, 013808 (2003).

\bibitem{exptransfer}T. A. Palomaki, J. W. Harlow, J. D. Teufel,
R. W. Simmonds, and K. W. Lehnert, Nature \textbf{495}, 210 (2013).

\bibitem{vg07} D. Vitali, S. Gigan, A. Ferreira, H. R. Bohm, P. Tombesi,
A. Guerreiro, V. Vedral, A. Zeilinger, and M. Aspelmeyer, Phys. Rev.
Lett. \textbf{98}, 030405 (2007).

\bibitem{gv08} C. Genes, D. Vitali, and P. Tombesi, Phys. Rev. A
\textbf{77}, 050307(R) (2008).

\bibitem{gb11} R. Ghobadi, A. R. Bahrampour, and C. Simon, Phys.
Rev. A \textbf{84}, 063827 (2011).

\bibitem{ab11} M. Abdi, Sh. Barzanjeh, P. Tombesi, and D. Vitali,
Phys. Rev. A \textbf{84}, 032325 (2011).

\bibitem{tgl11} H. T. Tan and G. X. Li, Phys. Rev. A \textbf{84},
024301 (2011).

\bibitem{am12} U. Akram, W. Munro, K. Nemoto, and G. J. Milburn,
Phys. Rev. A \textbf{86}, 042306 (2012).

\bibitem{jl12} C. Joshi, J. Larson, M. Jonson, E. Andersson, and
P. Ohberg, Phys. Rev. A \textbf{85}, 033805 (2012).

\bibitem{xz13} X. W. Xu, Y. J. Zhao, and Y. X. Liu, Phys. Rev. A
\textbf{88}, 022325 (2013).

\bibitem{sa14} P. Sekatski, M. Aspelmeyer, and N. Sangouard, Phys.
Rev. Lett. \textbf{112}, 080502 (2014).

\bibitem{tl13} H. Tan, G. X. Li, and P. Meystre, Phys. Rev. A \textbf{87},
033829 (2013).

\bibitem{aa14} M. Asjad, G. S. Agarwal, M. S. Kim, P. Tombesi, G.
Di Giuseppe, and D. Vitali, Phys. Rev. A \textbf{89}, 023849 (2014).

\bibitem{wc14} M. J. Woolley and A. A. Clerk, Phys. Rev. A \textbf{89},
063805 (2014).

\bibitem{ha11} S. Huang and G. S. Agarwal, Phys. Rev. A \textbf{83},
023823 (2011).

\bibitem{cs11} Y. Chang, T. Shi, Y. X. Liu, C. P. Sun, and F. Nori,
Phys. Rev. A \textbf{83}, 063826 (2011).

\bibitem{sn13} S. Shahidani, M. H. Naderi, and M. Soltanolkotabi,
Phys. Rev. A \textbf{88}, 053813 (2013).

\bibitem{exp_science2013}T. A. Palomaki, J. D. Teufel, R. W. Simmonds,
and K. W. Lehnert, Science \textbf{342}, 710 (2013).

\bibitem{2cavitymodes}M. C. Kuzyk, S. J. van Enk, and H. L. Wang,
Phys. Rev. A \textbf{88}, 062341 (2013).

\bibitem{ba11} Sh. Barzanjeh, D. Vitali, P. Tombesi, and G. J. Milburn,
Phys. Rev. A \textbf{84}, 042342 (2011).

\bibitem{cm11} G. De Chiara, M. Paternostro, and G. M. Palma, Phys.
Rev. A\textbf{ 83}, 052324 (2011). 

\bibitem{sl12} L. H. Sun, G. X. Li, and Z. Ficek, Phys. Rev. A \textbf{85},
022327 (2012). 

\bibitem{xb12} A. Xuereb, M. Barbieri, and M. Paternostro, Phys.
Rev. A\textbf{ 86}, 013809 (2012). 

\bibitem{rp12} B. Rogers, M. Paternostro, G. M. Palma, and G. De
Chiara, Phys. Rev. A\textbf{ 86}, 042323 (2012). 

\bibitem{wa13} Y. D. Wang and A. A. Clerk, Phys. Rev. Lett. \textbf{110},
253601 (2013). 

\bibitem{ti13} L. Tian, Phys. Rev. Lett. \textbf{110}, 233602 (2013). 

\bibitem{genu_ent_opto}Q. Y. He and Z. Ficek, Phys. Rev. A \textbf{89},
022332 (2014). 

\bibitem{s21} T. Holstein and H. Primakoff, Phys. Rev. \textbf{58},
1098 (1940). 

\bibitem{s20} A. S. Parkins, E. Solano, and J. I. Cirac, Phys. Rev.
Lett. \textbf{96}, 053602 (2006). 

\bibitem{Vitali2008}C. Genes, A. Mari, P. Tombesi, and D. Vitali,
Phys. Rev. A \textbf{78}, 032316 (2008).

\bibitem{gc85} C. W. Gardiner and M. J. Collett, Phys. Rev. A \textbf{31},
3761 (1985). 

\bibitem{mdr113} M. D. Reid, Phys. Rev. A \textbf{88}, 062108 (2013).

\bibitem{hb99} M. Hillery, V. Buzek, and A. Berthiaume, Phys. Rev.
A \textbf{59}, 1829 (1999). 

\bibitem{ag12} L. Aolita, R. Gallego, A. Cabello, and A. Acin, Phys.
Rev. Lett. \textbf{108}, 100401 (2012). 

\bibitem{lc14} Y. C. Liang, F. J. Curchod, J. Bowles, and N. Gisin,
arXiv:1405.3657 (2014).\end{thebibliography}
\end{document}